\documentclass[aps,prd,onecolumn,showpacs,nofootinbib,preprintnumbers,superscriptaddress]{revtex4}
\usepackage{bm}
\usepackage{amsmath,amssymb}
\usepackage{color}
\usepackage{graphicx}

\begin{document}

\title{Covariant St\"{u}ckelberg analysis of de Rham-Gabadadze-Tolley massive gravity with a general fiducial metric}

\author{Xian Gao}
\email[]{gao``at''th.phys.titech.ac.jp}
\affiliation{Department of Physics, Tokyo Institute of Technology, Tokyo 152-8551, Japan}

\author{Tsutomu Kobayashi}
\email[]{tsutomu``at''rikkyo.ac.jp}
\affiliation{Department of Physics, Rikkyo University, Toshima, Tokyo 175-8501, Japan}

\author{Masahide Yamaguchi}
\email[]{gucci``at''phys.titech.ac.jp}
\affiliation{Department of Physics, Tokyo Institute of Technology, Tokyo 152-8551, Japan}

\author{Daisuke Yoshida}
\email[]{yoshida``at''th.phys.titech.ac.jp}
\affiliation{Department of Physics, Tokyo Institute of Technology, Tokyo 152-8551, Japan}

\begin{abstract}
The St\"{u}ckelberg analysis of nonlinear massive gravity in the
presence of a general fiducial metric is investigated.  We develop a
 ``covariant'' formalism for the St\"{u}ckelberg expansion by working
with a local inertial frame, through which helicity modes can be
characterized correctly.  Within this covariant approach, an extended
$\Lambda_3$ decoupling limit analysis can be consistently performed,
which keeps $\bar{R}_{\mu\nu\rho\sigma}/m^2$ fixed with
$\bar{R}_{\mu\nu\rho\sigma}$ the Riemann tensor of the fiducial metric.
In this extended decoupling limit, the scalar mode $\pi$ acquires
self-interactions due to the presence of the curvature of the fiducial
metric.  However, the equation of motion for $\pi$ remains of second
order in derivatives, which extends the understanding of the absence of
the Boulware Deser ghost in the case of a flat fiducial metric.
\end{abstract}

\preprint{RUP-14-13}

\maketitle

\section{Introduction}
Massive gravity is a candidate of modified gravity which explains the
current accelerated expansion of the Universe~\cite{Riess:1998cb,
Perlmutter:1998np}.  A linear theory of massive graviton was first
considered by Fierz and Pauli (FP)~\cite{Fierz:1939ix}. This theory
succeeded in excluding the extra ghost degree of freedom at linear
order, but van Dam, Veltman, and Zakharov suggested that this theory
does not reduce to general relativity (GR) even in the massless
limit~\cite{vanDam:1970vg,Zakharov:1970cc}.  However, Vainshtein showed
that this problem is caused by omitting the nonlinear
effects~\cite{Vainshtein:1972sx}. For this reason, nonlinear extensions
of the FP theory have been actively considered. However, for a long
time, these nonlinear theories had suffered from the Boulware Deser (BD)
ghost problem~\cite{Boulware:1973my}, which states that the theories
have the extra ghost degree of freedom in addition to the usual 5
degrees of freedom of massive spin-2.
 
In order to see the BD ghost explicitly, the St\"{u}ckelberg formalism is
very useful~\cite{ArkaniHamed:2002sp}. In St\"{u}ckelberg language, the
physical degrees of freedom are decomposed into helicity-0, helicity-1, and
helicity-2 modes. The origin of the BD ghost mode is understood as the
higher order equation of motion for the helicity-0 mode.
In Refs.~\cite{deRham:2010ik,deRham:2010kj}, de Rham, Gabadadze, and
Tolley (dRGT) constructed the mass potential
in which the self-interactions of helicity-0 mode is tuned to be a
total divergence, leaving a second-order equation of motion for the
helicity-0 mode. It can be seen that after a field redefinition the
action for the helicity-0 mode indeed reduces to the Galileon
form~\cite{Nicolis:2008in} clearly having the second-order equation of
motion. Thus, this theory is BD ghost free. Massive gravity with this
mass potential is called the dRGT theory.
Hassan and Rosen investigated dRGT massive gravity by means of the Hamiltonian analysis,
and proved that the dRGT theory is free of the BD ghost even away from
the decoupling limit~\cite{Hassan:2011hr,Hassan:2011ea}. In addition, they generalized
dRGT massive gravity on a flat fiducial metric to the theory on a
general fiducial metric, and proved that the theory is also free of the
BD ghost using the same method~\cite{Hassan:2011tf,Hassan:2011ea}.

Although formal Hamiltonian analysis is sufficient to prove the absence of ghost, it is interesting to show this explicitly at the level of equations of motion in a St\"{u}ckelberg language.
In a complementary way to the Hamiltonian analysis of Hassan and Rosen,
the St\"{u}ckelberg analysis in the flat fiducial case has been studied
in detail in
Refs.~\cite{deRham:2011rn,Mirbabayi:2011aa,Ondo:2013wka,Kugo:2014hja}.
Moreover, the St\"{u}ckelberg analysis was extended to the curved
fiducial case: de Rham and Renaux-Petel studied the de Sitter fiducial
case~\cite{deRham:2012kf}, while Fasiello and Tolley analyzed the
Friedmann-Lemaitre-Robertson-Walker fiducial
case~\cite{Fasiello:2013woa}.  However, the St\"{u}ckelberg analysis has
not been performed so far in the general fiducial case. 
(Hamiltonian analysis in St\"{u}ckelberg language is discussed in Ref.~\cite{Hassan:2012qv}.)
In this paper,
we discuss the St\"{u}ckelberg analysis in the dRGT theory with a
completely general fiducial metric. First, we extend the definition of
the perturbation of the St\"{u}ckelberg field in a covariant
manner. Using this definition, we expand the action in terms of the
perturbed quantities up to fourth order. Next, we extend the decoupling
limit of the flat case to that of a curved one by scaling the curvature
scale. Finally, we show that, in this extended decoupling limit, the
equation of motion for the helicity-0 mode $\pi$ does not include higher
derivatives of $\pi$, which clarifies in a different and complementary
way the reason why the BD ghost is absent even in the curved fiducial
case.

This paper is organized as follows. In the next section, we briefly
review dRGT massive gravity and the usual St\"{u}ckelberg
analysis. In Sec. \ref{sec:cov}, we develop a covariant St\"{u}ckelberg
analysis with a general fiducial metric and apply it to dRGT
massive gravity. In Sec.\ref{sec:dec}, we derive the decoupling limit at
the energy scale $\Lambda_3$ and show that the equation of motion for
$\pi$ remains of second order in derivatives. The final section is
devoted to the conclusions.

\section{Massive gravity with a curved fiducial metric}
The Lagrangian for nonlinear massive gravity is composed of the
Einstein-Hilbert term and the dRGT mass terms:
\begin{equation}
		S=\frac{M_{\mathrm{pl}}^{2}}{2}\int d^{4}x\sqrt{-g}\left(R+m^{2}\mathcal{L}^{\mathrm{dRGT}}\right), \label{dRGT}
\end{equation}
where $m$ is the mass of the graviton $h_{\mu\nu}$ defined by
\begin{equation}
		 h_{\mu\nu} \equiv g_{\mu\nu}- {g}^{(0)}_{\mu\nu},
\end{equation}
with a fixed background metric ${g}^{(0)}_{\mu\nu}$. The dRGT mass terms
are given by
\begin{equation}
		\mathcal{L}^{\mathrm{dRGT}}=\mathcal{U}_{2}+\alpha_{3}\, \mathcal{U}_{3}+\alpha_4 \,\mathcal{U}_{4}, \label{dRGT_mass}
\end{equation}
where
\begin{eqnarray}
		{\cal U}_2 &=& [{\cal K}]^2 - [{\cal K}^2], \label{U2}\\ 
		{\cal U}_3 &=& [{\cal K}]^3 -3[{\cal K}][{\cal K}^2]+2 [{\cal K}^3], \label{U3}\\ 
		{\cal U}_4 &=& [{\cal K}]^4-6[{\cal K}]^2[{\cal K}^2]+8[{\cal K}][{\cal K}^3]+3[{\cal K}^2]^2 - 6[{\cal K}^4], \label{U4}
\end{eqnarray}
with $\alpha_3$ and $\alpha_4$ being free constant parameters. The
square brackets ``$[{\cal M}]$'' denote the trace of the matrix ${\cal M}$ with respect to
the physical metric $g_{\mu\nu}$. The matrix
 ${\cal K}^{\mu}_{\phantom{\mu}\nu}$ is defined by
	\begin{equation}
	 {\cal K}^\mu_{\ \nu} = \delta^\mu_{\ \nu}-\sqrt{g^{\mu\rho}\bar{g}_{\rho\nu}}, \label{K_def}
	\end{equation}
where $\bar{g}_{\mu\nu}$ is a fixed symmetric matrix that is called a
fiducial metric. Generally speaking, a background metric
$g^{(0)}_{\mu\nu}$ on which we define a graviton has nothing to do with
the fiducial metric $\bar{g}_{\mu\nu}$, though in many cases
$\bar{g}_{\mu\nu}$ is indeed a solution of the background equation of
motion.  In this paper, we take $g^{(0)}_{\mu\nu}= \bar{g}_{\mu\nu}$
for simplicity, which enables us to expand the matrix ${\cal K}$
perturbatively.

\subsection{St\"{u}ckelberg trick}

The matrix ${\cal K}^{\mu}_{\phantom{\mu}\nu}$ and thus the dRGT mass
terms (\ref{U2})--(\ref{U4}) explicitly break general covariance due to
the presence of a fixed fiducial metric $\bar{g}_{\mu\nu}$.  On the
other hand, $\bar{g}_{\mu\nu}$ can always be thought of as the
``gauge-fixed'' version of some covariant tensor field, which can be
constructed using the well-known St\"{u}ckelberg trick
\cite{ArkaniHamed:2002sp,Siegel:1993sk,Creminelli:2005qk}:
	\begin{eqnarray}
		\bar{g}_{\mu\nu}(x)\rightarrow f_{\mu\nu}(x)=\bar{g}_{ab}(\phi(x))\frac{\partial\phi^{a}(x)}{\partial x^{\mu}}\frac{\partial\phi^{b}(x)}{\partial x^{\nu}}, \label{fiducial}
	\end{eqnarray}
where a set of four (we are working in four-dimensional spacetime)
St\"{u}ckelberg fields $\{\phi^a\}$ transform as scalars under a general
coordinate transformation of spacetime. The fixed $\bar{g}_{\mu\nu}$ can
be recovered by choosing the so-called ``unitary gauge'' with $\phi^\mu
= x^\mu$.  By replacing $\bar{g}_{\mu\nu}$ with the covariant tensor
field $f_{\mu\nu}$, the dRGT mass terms are promoted to scalars and thus the
corresponding Lagrangian (\ref{dRGT}) acquires general covariance.

Degrees of freedom in a gravity theory alternative to GR show themselves
in a simpler manner in the so-called ``decoupling limit,'' where
different types (e.g., helicities) of degrees of freedom decouple from
each other in some limit of energy scales.  In the case of massive
gravity, the decoupling limit is taken as $M_{\mathrm{pl}} \rightarrow
\infty$ so that the nonlinearities in gravity get reduced, while keeping
the energy scale $\Lambda_{\lambda}\equiv \left(M_\mathrm{pl}
m^{\lambda-1}\right)^{\lambda}$ with some $\lambda$ fixed so that
interactions arising above $\Lambda_{\lambda}$ become irrelevant.  When
a fiducial metric is flat, all degrees of freedom are thus living in the
flat Minkowski background, which enables us to identify the
(St\"{u}ckelberg) field-space Lorentz symmetry with the spacetime global
Lorentz symmetry, while the later only arises in the decoupling limit.
In this case, the set of four St\"{u}ckelberg fields $\{\phi^{a}\}$
transform as a vector under this identified global Lorentz
transformation. As a result, the fields $A_{a}$ and $\pi$, defined
by
	\begin{equation}
		\phi^a = x^a - \pi^a =  x^a - A^a - \partial^a \pi, \label{split}
	\end{equation}
transform as a vector and a scalar and encode the information of
helicity-1 and helicity-0 modes of a massive graviton, respectively, in
the decoupling limit. Thus, in this limit, the requirement
that the equation of motion for $\pi$ is of second order in derivatives
ensures the absence of BD ghost~\cite{Deffayet:2005ys,Creminelli:2005qk}.

The above argument, however, cannot be applied simply to the case of a
curved fiducial metric. Indeed, a naive split~(\ref{split}) in
Eq.~(\ref{fiducial}) yields
	\begin{equation}
		f_{\mu\nu}=\bar{g}_{\mu\nu}-\bar{g}_{\rho\mu}\partial_{\nu}\pi^{\rho}-\bar{g}_{\rho\nu}\partial_{\mu}\pi^{\rho}+\bar{g}_{\rho\sigma}\partial_{\mu}\pi^{\rho}\partial_{\nu}\pi^{\sigma},
	\end{equation}
where $\pi^{\mu}$ is defined as $\pi^a \equiv \delta^a_{\mu}
\pi^{\mu}$.  If we would rewrite all partial derivatives in terms of
``covariant'' derivatives $\bar{\nabla}$ with respect to
$\bar{g}_{\mu\nu}$ using
$\partial_{\mu}\pi^{\rho}=\bar{\nabla}_{\mu}\pi^{\rho}-\bar{\Gamma}_{\mu\lambda}^{\rho}\pi^{\lambda}$
and focus on the ``supposed-to-be'' helicity-0 mode $\pi$ defined
by $\pi^{\mu}\equiv\bar{g}^{\mu\nu}\partial_{\nu}\pi$, we will get
	\begin{eqnarray}
	f_{\mu\nu} & = & \bar{g}_{\mu\nu}-2\bar{\nabla}_{\mu}\bar{\nabla}_{\nu}\pi+\bar{\nabla}_{\mu}\bar{\nabla}_{\sigma}\pi\bar{\nabla}_{\nu}\bar{\nabla}^{\sigma}\pi\nonumber \\
	 &  & +\left(\bar{g}_{\rho\mu}\bar{\Gamma}_{\nu\lambda}^{\rho}+\bar{g}_{\rho\nu}\bar{\Gamma}_{\mu\lambda}^{\rho}\right)\bar{\nabla}^{\lambda}\pi\nonumber \\
	 &  & -\bar{\nabla}_{\mu}\bar{\nabla}_{\sigma}\pi\bar{\Gamma}_{\nu\tau}^{\sigma}\bar{\nabla}^{\tau}\pi-\bar{\nabla}_{\nu}\bar{\nabla}_{\rho}\pi\bar{\Gamma}_{\mu\lambda}^{\rho}\bar{\nabla}^{\lambda}\pi+\bar{g}_{\rho\sigma}\bar{\Gamma}_{\mu\lambda}^{\rho}\bar{\Gamma}_{\nu\tau}^{\sigma}\bar{\nabla}^{\lambda}\pi\bar{\nabla}^{\tau}\pi,
	\end{eqnarray}
which is not explicitly covariant due to the presence of the Christoffel
symbols.

In fact, as was well explained in  \cite{Hassan:2011vm} (see
also \cite{deRham:2011rn,deRham:2012kf}), $A_{a}$ and $\pi$ defined in the ``naive''
split (\ref{split}) are neither vector nor scalar any longer, and do not
capture the helicity-1 and helicity-0 modes correctly, either when going
beyond the decoupling limit or when the fiducial metric is curved, which
is the case we are dealing with in this paper.

The main purpose of this paper is thus to develop a covariant
formalism for the St\"{u}ckelberg expansion, through which the helicity
modes can be characterized correctly and a decoupling limit analysis
similar to the case of a flat fiducial metric can be consistently
performed.

\section{A covariant approach to St\"{u}ckelberg analysis} \label{sec:cov}

\subsection{Covariant definition of the perturbed St\"{u}ckelberg field}
\label{CovDefpi}

For a general fiducial metric, the main difficulty in ``covariantly''
defining the Goldstone modes $\pi^{\mu}$ and identifying the helicity
modes is that, due to the loss of global symmetries, the identification
of the internal and physical space does not make sense any longer.  One
exception is the case of the maximally symmetric fiducial metric
considered in \cite{deRham:2012kf}, where the identification of the
helicity modes was made by embedding the $d$-dimensional (A)dS into a
$(d+1)$-dimensional Minkowski background and then projecting back. This
trick, however, cannot be used for a general fiducial metric since
embedding an arbitrary $d$-dimensional space into a $(d+1)$-dimensional
Minkowski one is not always possible.

In this paper, we employ an alternative approach based on the Riemann
normal coordinates (RNC), which is in fact a standard approach to
defining perturbations covariantly, as has been used in the
well-known background field method
(e.g.~\cite{AlvarezGaume:1981hn}). See also footnote 5 of
\cite{Hassan:2011vm}, in which the use of Riemann normal coordinate is
suggested. The idea is to regard the St\"{u}ckelberg field as the
diffeomorphism of the spacetime itself.\footnote{See also Appendix
\ref{sc:fieldspace} for an alternative point of view of the field space,
which yields exactly the same definition for the Goldstone modes
$\pi^{\mu}$ as in Eq.~(\ref{expandphipi}).}  
Precisely, we consider a
one-parameter family of diffeomorphisms generated by a set of
single-parameter curves $x^{\mu}(\lambda)$ parametrized by $\lambda$,
i.e.,
	\begin{equation}
		\phi_{\lambda}:\quad p \mapsto \phi_{\lambda}(p),
	\end{equation}
for a given point $p$ in spacetime.
At this point, we do not assume $x^\mu(\lambda)$ to be a geodesic, while we shall see below how the standard RNC approach arises in order to recover the expressions in the case of a flat fiducial metric (\ref{split}).
We may freely set $\lambda=0$ at a given point $p$, and define the St\"{u}ckelberg fields at $p$ as the coordinate values of its image $\phi_{\lambda}(p)$ at $\lambda = -1$, i.e.,
	\begin{equation}
		\left.\phi^{\mu}\right|_{p}\equiv\left.x^{\mu}\right|_{\phi_{-1}\left(p\right)}=x^{\mu}(-1). \label{stu_def}
	\end{equation}
Note that in Eq.~(\ref{stu_def})
we use the same symbol both for the diffeomorphism and for the St\"{u}ckelberg fields.
The perturbation of the St\"{u}ckelberg fields, i.e.,
the difference between the St\"{u}ckelberg fields at point $p$ and the coordinate values of point $p$ itself,
	\begin{equation}
		\left.\phi^{\mu}\right|_{p}-\left.x^{\mu}\right|_{p}=x^{\mu}(-1)-x^{\mu}(0),
	\end{equation}
is obviously not a covariant object, since $x^{\mu}(0)$ are fixed.

\begin{figure}[t!]
	\centering
	\begin{minipage}{0.8\textwidth}
	\includegraphics[width=10cm]{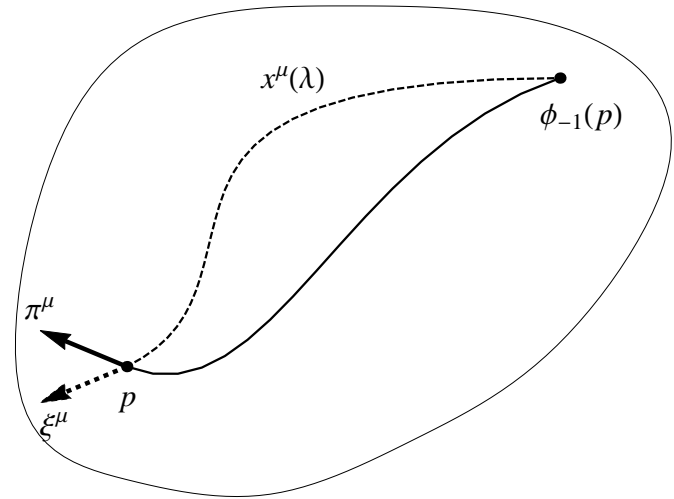}
	\caption{
	Illustration of the definitions of St\"{u}ckelberg field $\phi^{\mu}$ and the Goldstone modes $\pi^{\mu}$. 
	The St\"{u}ckelberg fields $\left.\phi^{\mu}\right|_p$ at a given point $p$ are defined as the coordinate values $\left.x^{\mu}\right|_{\phi_{-1}(p)}$, where $\phi_{-1}(p)$ is the image of $p$ under the diffeomorphism generated by single-parameter curves $x^{\mu}(\lambda)$ with parameter $\lambda=-1$.
	If $x^{\mu}(\lambda)$ are not geodesics (dashed curve), the St\"{u}ckelberg field is expanded as (\ref{phi_expan}); if $x^{\mu}(\lambda)$ are geodesics (black curve) the St\"{u}ckelberg field is expanded as (\ref{expandphipi}) with Goldstone modes $\pi^{\mu}$ as the standard Riemann normal coordinates.
	}
		\label{fig:map}
	\end{minipage}
	\end{figure}

We define $u^\mu$ as the tangent
vector of $x^{\mu}(\lambda)$,
	\begin{equation}
		u^{\mu}(\lambda)\equiv \frac{\mathrm{d}x^{\mu}(\lambda)}{\mathrm{d} \lambda}, \label{defpi} 
	\end{equation}
which is automatically a covariant object by definition.
Integrating Eq.~(\ref{defpi}) gives
	\begin{eqnarray}
	x^{\mu}\left(\lambda\right) & \equiv  & \left.\left[e^{\lambda\frac{\mathrm{d} }{\mathrm{d}\lambda'}}x^{\mu}\left(\lambda'\right)\right]\right|_{\lambda'=0}\nonumber \\
	 & = & \left.\left[\exp\left(\lambda u^{\alpha}\left(\lambda'\right)\frac{\partial}{\partial x^{\alpha}\left(\lambda'\right)}\right)x^{\mu}\left(\lambda'\right)\right]\right|_{\lambda'=0}\nonumber \\
	 & = & x_{0}^{\mu}+\lambda\xi^{\mu}+\frac{\lambda^{2}}{2}\xi^{\alpha}\left.\frac{\partial u^{\mu}\left(\lambda'\right)}{\partial x^{\alpha}}\right|_{\lambda'=0}+\frac{\lambda^{3}}{3!}\xi^{\alpha}\left.\frac{\partial}{\partial x^{\alpha}}\left(u^{\beta}\left(\lambda'\right)\frac{\partial u^{\mu}\left(\lambda'\right)}{\partial x^{\beta}}\right)\right|_{\lambda'=0}+\cdots, \label{exponential}
	\end{eqnarray}
where we used $\frac{\mathrm{d}}{\mathrm{d}\lambda}\equiv u^{\alpha}\left(\lambda\right)\frac{\partial}{\partial x^{\alpha}\left(\lambda\right)}$ and denoted 
	\begin{equation}
		\xi^\mu \equiv u^\mu(0), \qquad  x^\mu_0 \equiv x^\mu(0),
	\end{equation}
for short.
Setting $\lambda=-1$ in Eq.~(\ref{exponential}) yields
	\begin{equation}
		\left.\phi^{\mu}\right|_p \equiv x^{\mu}\left(-1\right)= x_{0}^{\mu}-\xi^{\mu}+\frac{1}{2}\xi^{\alpha}\partial_{\alpha}\xi^{\mu}-\frac{1}{3!}\xi^{\alpha}\partial_{\alpha}\left(\xi^{\beta}\partial_{\beta}\xi^{\mu}\right)+\mathcal{O}(\xi^4). \label{phi_expan}
	\end{equation}
Note that Eq.~(\ref{exponential}) and thus Eq.~(\ref{phi_expan})
are the results of the
standard Taylor expansion and we do not assume that
$x^{\mu}(\lambda)$ is a geodesic.
If we naively identify $\xi^{\mu}$ as the Goldstone modes, due to the presence
of derivatives of $\xi^{\mu}$, (\ref{phi_expan}) does not reduce to Eq.~(\ref{split})
even in the case of the flat fiducial metric.
This discrepancy can be trivially solved by introducing a new variable $\pi^{\mu}$ through
	\begin{equation}
		\pi^{\mu}=\xi^{\mu}-\frac{1}{2}\xi^{\alpha}\bar{\nabla}_{\alpha}\xi^{\mu}+\frac{1}{3!}\xi^{\alpha}\bar{\nabla}_{\alpha}\left(\xi^{\beta}\bar{\nabla}_{\beta}\xi^{\mu}\right)+\mathcal{O}\left(\xi^{4}\right), \label{pi_def}
	\end{equation}
or its inverted form
	\begin{equation}
		\xi^{\mu}=\pi^{\mu}+\frac{1}{2}a^{\mu}+\frac{1}{12}\pi^{\rho}\bar{\nabla}_{\rho}a^{\mu}+\frac{1}{4}a^{\nu}\bar{\nabla}_{\nu}\pi^{\mu}+\mathcal{O}\left(\pi^{4}\right), \label{xi_pi}
	\end{equation}
where $\bar{\nabla}_\mu$ is the covariant derivative with respect to
$\bar{g}_{\mu\nu}$ and $a^\mu \equiv \pi^\nu\bar{\nabla}_\nu \pi^\mu$.
Since $\xi^{\mu}$ is covariant, $\pi^{\mu}$ defined by
Eq.~(\ref{pi_def}) is automatically covariant, which we identify as the
covariant Goldstone modes.  Plugging Eq.~(\ref{xi_pi}) into
Eq.~(\ref{phi_expan}) and performing some manipulation, we have
	\begin{equation}
		\left.\phi^\mu\right|_p =x^\mu_0-\pi^\mu-\frac{1}{2}\bar{\Gamma}^{\mu}_{\nu\rho}\pi^\nu\pi^\rho
		+\frac{1}{6}
		\left(
		\partial_\nu\bar{\Gamma}^{\mu}_{\rho\sigma}-2\bar{\Gamma}^{\mu}_{\nu\lambda}\bar{\Gamma}^{\lambda}_{\rho\sigma}
		\right)
		\pi^\nu\pi^\rho\pi^\sigma+\cdots,\label{expandphipi}
	\end{equation}
which indeed
reduces to Eq.~(\ref{split}) in the case of the flat fiducial metric.
Equation~(\ref{expandphipi})
is one of the main results of this paper.
This result implies that
the covariant Goldstone modes $\pi^{\mu}$ are nothing but the standard Riemann normal coordinates, which  correspond to the tangent vector of the geodesic at point $p$ connecting the point $p$ and its image $\phi_{-1}(p)$.
The definitions of the
St\"{u}ckelberg field $\phi^{\mu}$ and the Goldstone modes  $\pi^{\mu}$ are illustrated in
Fig.~\ref{fig:map}.

Before ending this subsection, we emphasize again that both $\xi^{\mu}$ and $\pi^{\mu}$ are covariant and have explicit covariant relations (\ref{pi_def}) or (\ref{xi_pi}), and thus both
Eqs.~(\ref{phi_expan}) and~(\ref{expandphipi}) can be used to
derive covariant expressions. We employ Eq.~(\ref{expandphipi})
since this has an explicit correspondence to the case of the flat fiducial metric (\ref{split}).

\subsection{St\"{u}ckelberg expansion of the action}

Having defined the perturbative expansion of the St\"{u}ckelberg and the Goldstone modes $\pi^{\mu}$
as in Eq.~(\ref{expandphipi}), we are able to expand the ``covariantized'' fiducial metric 
$f_{\mu\nu}$ in Eq.~(\ref{fiducial}). 
Simply by plugging Eq.~(\ref{expandphipi}) and
carefully dealing with the Christoffel symbols and their derivatives, it
is straightforward to recast the expressions in terms of $\pi^{\mu}$, $\bar{R}_{\mu\nu\rho\sigma}$,
and their covariant derivatives (with respect to the fiducial metric). This procedure, however, becomes more and more cumbersome when going to higher orders in $\pi^{\mu}$.

In this subsection, we take an equivalent but simpler treatment by recalling that, according to
Eq.~(\ref{fiducial}),
$f_{\mu\nu}$ can be viewed as the ``pull back'' of $\bar{g}_{\mu\nu}$ under the diffeomorphism $\phi$ of spacetime:
	\begin{equation}
		\left.f_{\mu\nu}\right|_{p}=\left.\left(\phi_{-1}^{\ast}\,\bar{g}_{\mu\nu}\right)\right|_{p},
	\end{equation}
for a given point $p$.
As in the standard lore, an infinitesimal diffeomorphism is generated by a vector field, which is just the tangent vector of the curve $u^{\mu} = \mathrm{d}x^{\mu}/\mathrm{d}\lambda$ in our case.
The change in any tensor field induced by such an infinitesimal diffeomorphism is encoded in the Lie derivatives along the curve $x^{\mu}(\lambda)$. 
Precisely, we introduce a coordinate system adapted to $u^{\mu}$, such that 
the parameter $\lambda$ along the curve $x^{\mu}(\lambda)$ is chosen as one of the coordinates, e.g., $x^0$, while the values of other coordinates $\{x^i\}$ are kept invariant along the curve.
In this particular coordinate system, for a given point $p$ with coordinate values $x^\mu\equiv (x^0,x^i)$, we have $\left.\phi^{\mu}\right|_{p}=(x^{0}-1,x^{i})$ and thus
	\begin{eqnarray}
	\left.\frac{\partial\phi^{\mu}}{\partial x^{\nu}}\right|_{p}=\delta_{\phantom{\mu}\nu}^{\mu}.\label{deltaamu}
	\end{eqnarray}
The components of $f_{\mu\nu}$ in this peculiar coordinate system are thus given by 
	\begin{eqnarray}
	f_{\mu\nu}\left(x\right)\equiv\left.f_{\mu\nu}\right|_{p} & = & \left.\left(\phi_{-1}^{\ast}\bar{g}_{\mu\nu}\right)\right|_{p}\nonumber \\
	 & = & \bar{g}_{\mu\nu}\left(x^{0}-1,x^{i}\right)\nonumber \\
	 & \equiv & e^{-\partial/\partial x^{0}}\bar{g}_{\mu\nu}\left(x\right).\label{f_exp}
	\end{eqnarray}
On the other hand, in this peculiar coordinate system, $\partial/\partial x^0 $ is equivalent to the Lie derivative $\pounds_{u}$ when acting on any tensor,
Eq.~(\ref{f_exp}) can be recast into a covariant form\footnote{The same expression
is used in cosmological perturbations of massive gravity around
an open-Friedmann-Robertson-Walker fiducial metric~\cite{Gumrukcuoglu:2011zh}. 
Similar higher order Lie derivatives also naturally arise when constructing higher-order
gauge-invariant cosmological perturbations, see e.g.~\cite{Bruni:1996im}.}:
	\begin{equation}
		f_{\mu\nu}(x)=e^{-\pounds_{\xi}}\bar{g}_{\mu\nu}(x), \label{f_exp_lie}
	\end{equation}
where $\xi^{\mu} \equiv \left.u^{\mu}\right|_p$.
Equation~(\ref{f_exp_lie}) is also one of the main results of this
paper, which now actually holds in any coordinate system.

Using Eq.~(\ref{f_exp_lie}), it is now straightforward to expand $f_{\mu\nu}$ in term of
$\xi^\mu$:
	\begin{eqnarray}
	f_{\mu\nu} & = & \bar{g}_{\mu\nu}-2\bar{\nabla}_{(\mu}\xi_{\nu)}\nonumber \\
	 &  & +\bar{\nabla}_{\mu}\xi_{\rho}\bar{\nabla}_{\nu}\xi^{\rho}-\bar{R}_{\mu\rho\nu\sigma}\xi^{\rho}\xi^{\sigma}+\bar{\nabla}_{(\mu}\tilde{a}_{\nu)}\nonumber \\
	 &  & +\frac{1}{3}\bar{\nabla}_{\lambda}\bar{R}_{\mu\rho\nu\sigma}\xi^{\lambda}\xi^{\rho}\xi^{\sigma}+\frac{2}{3}\bar{R}_{\mu\rho\lambda\sigma}\bar{\nabla}_{\nu}\xi^{\lambda}\xi^{\rho}\xi^{\sigma}+\frac{2}{3}\bar{R}_{\nu\rho\lambda\sigma}\bar{\nabla}_{\mu}\xi^{\lambda}\xi^{\rho}\xi^{\sigma}\nonumber \\
	 &  & +\bar{R}_{(\mu|\rho|\nu)\sigma}\tilde{a}^{\rho}\xi^{\sigma}-\bar{\nabla}_{(\mu}\xi^{\rho}\bar{\nabla}_{\nu)}\tilde{a}_{\rho}-\frac{1}{3}\bar{\nabla}_{(\mu|}\left(\xi^{\rho}\nabla_{\rho}\tilde{a}_{|\nu)}\right)+{\cal O}(\xi^{4}),\label{f_expan}
	\end{eqnarray}
where all indices are raised and lowered by $\bar{g}^{\mu\nu}$ and
$\bar{g}_{\mu\nu}$, respectively, $\bar{R}^{\mu}_{\ \nu\rho\sigma}$ is
the Riemann tensor constructed from $\bar{g}_{\mu\nu}$, and
$\tilde{a}^{\mu} \equiv \xi^{\nu}\bar{\nabla}_{\nu}\xi^{\mu}$.  It is
more convenient to eliminate $\tilde{a}^{\mu}$ by using $\pi^{\mu}$
defined by Eq.~(\ref{pi_def}) or Eq.~(\ref{xi_pi}), which yields
	\begin{eqnarray}
	\tilde{a}^{\mu} & \equiv & \xi^{\nu}\bar{\nabla}_{\nu}\xi^{\mu}\nonumber \\
	 & = & a^{\mu}+\frac{1}{2}\pi^{\nu}\bar{\nabla}_{\nu}a^{\mu}+\frac{1}{2}a^{\nu}\bar{\nabla}_{\nu}\pi^{\mu}+{\cal O}(\pi^{4}),\label{a_tilde}
	\end{eqnarray}
with $a^{\mu} \equiv \pi^{\nu} \bar{\nabla}_{\nu} \pi^{\mu}$.
Plugging Eq.~(\ref{a_tilde}) into Eq.~(\ref{f_expan})
and performing simple manipulations, we  get
	\begin{eqnarray}
		f_{\mu\nu}=
		f^{(0)}_{\mu\nu}+
		f^{(1)}_{\mu\nu}+
		f^{(2)}_{\mu\nu}+
		f^{(3)}_{\mu\nu}+{\cal O}(\pi^4),
	\end{eqnarray}
with
	\begin{eqnarray}
		f_{\mu\nu}^{(0)} &=& \bar{g}_{\mu\nu}, \label{f_exp_0}\\
		f_{\mu\nu}^{(1)} &=& -2 \bar{\nabla}_{(\mu} \pi_{\nu)}, \label{f_exp_1}\\
		f_{\mu\nu}^{(2)} &=&\bar{\nabla}_\mu \pi_\rho \bar{\nabla}_\nu \pi^\rho-\bar{R}_{\mu\rho\nu\sigma}\pi^\rho \pi^\sigma
		,\label{f_exp_2}\\
		f^{(3)}_{\mu\nu}
		&=&\frac{1}{3}\bar{\nabla}_\lambda\bar{R}_{\mu\rho\nu\sigma}\pi^\lambda \pi^\rho \pi^\sigma
		+\frac{2}{3}\bar{R}_{\mu\rho\lambda\sigma}\bar{\nabla}_\nu \pi^\lambda \pi^\rho \pi^\sigma
		+
		\frac{2}{3}\bar{R}_{\nu\rho\lambda\sigma}\bar{\nabla}_\mu \pi^\lambda \pi^\rho \pi^\sigma.\label{f_exp_3}
	\end{eqnarray}
It is not surprising that the acceleration $a^{\mu}$ drops out in the
above expression. In fact, Eqs.~(\ref{f_exp_0})--(\ref{f_exp_3}) can
also be derived by replacing $\xi^{\mu}$ by $\pi^{\mu}$ in
Eq.~(\ref{f_exp_lie}) and taking into account that $a^{\mu}\equiv 0$
(since $\pi^{\mu}$ is the tangent vector of geodesics) when evaluating
the Lie derivatives.  This is also one of the advantages of defining the
Goldstone modes $\pi^{\mu}$ using the standard Riemann normal
coordinates.

Having the above results in hand, we are now ready to expand the covariant metric perturbation 
$H_{\mu\nu}\equiv g_{\mu\nu}-f_{\mu\nu}$ as
\begin{eqnarray}
H_{\mu\nu}=H^{(1)}_{\mu\nu}+H^{(2)}_{\mu\nu}+H^{(3)}_{\mu\nu}+{\cal O}(\pi^4), \label{H_exp}
\end{eqnarray}
with
	\begin{eqnarray}
		H_{\mu\nu}^{(1)} &=&h_{\mu\nu}+2 \bar{\nabla}_{(\mu} \pi_{\nu)}, \label{H_exp_1}\\
		H_{\mu\nu}^{(2)} &=&-\bar{\nabla}_\mu \pi_\rho \bar{\nabla}_\nu\pi^\rho+\bar{R}_{\mu\rho\nu\sigma}\pi^\rho\pi^\sigma,\label{H_exp_2}\\
		H^{(3)}_{\mu\nu} &=& -\frac{1}{3}\bar{\nabla}_\lambda\bar{R}_{\mu\rho\nu\sigma}\pi^\lambda\pi^\rho \pi^\sigma
		-\frac{2}{3}\bar{R}_{\mu\rho\lambda\sigma}\bar{\nabla}_\nu\pi^\lambda\pi^\rho \pi^\sigma
		-
		\frac{2}{3}\bar{R}_{\nu\rho\lambda\sigma}\bar{\nabla}_\mu\pi^\lambda\pi^\rho \pi^\sigma. \label{H_exp_3}
	\end{eqnarray}
In order to expand the matrix ${\cal K}^\mu_{\ \nu}$
in terms of $\pi^{\mu}$, first we note, by definition (\ref{K_def}), that
	\begin{eqnarray}
		{\cal K}^\mu_{\ \nu}&\equiv &\delta^\mu_{\ \nu}
		-\sqrt{g^{\mu\rho}f_{\rho\nu}}\notag\\
		&=&\delta^\mu_{\ \nu}
		-\sqrt{\delta^{\mu}_{\ \nu}-(g^{-1}H)^\mu_{\ \nu}}\notag\\
		&=&
		\frac{1}{2}(g^{-1}H)^\mu_{\ \nu}+\frac{1}{8}(g^{-1}H)^{2\mu}_{\ \ \nu}+\frac{1}{16}(g^{-1}H)^{3\mu}_{\ \ \nu}+\cdots,\label{calK}
	\end{eqnarray}
with $(g^{-1}H)^\mu_{\ \nu} \equiv g^{\mu\rho}H_{\rho\nu}$ etc.  When
$g_{\mu\nu}$ and $f_{\mu\nu}$ are expanded around the same background
(as we will do in this paper), $H_{\mu\nu}$ is a perturbative quantity
and hence we can calculate the action order by order through this
equation.\footnote{Note that the square root matrix can be also expanded
easily in the proportional background case:
$\bar{g}_{\mu\nu}=C^2g^{(0)}_{\mu\nu}$. Our following analysis can be
easily extended to this case. The only difference is existence of the overall
factor $C$ and ${\cal K}^{(0)}{}^{\mu}_{\nu} =
(1-C)\delta^{\mu}{}_{\nu}$.}
Plugging
Eqs.~(\ref{H_exp_1})--(\ref{H_exp_3}) into Eq.~(\ref{calK}) and using
$g^{\mu\nu} = \bar{g}^{\mu\nu}-h^{\mu\nu}+h^{\mu}_{\
\rho}h^{\rho\nu}+\cdots$, ${\cal K}^\mu_{\ \nu}$ can be expanded as
	\begin{eqnarray}
		{\cal K}^\mu_{\ \nu} = {\cal K}^{(1)}{}^\mu_{\ \nu}+{\cal K}^{(2)}{}^\mu_{\ \nu}+{\cal K}^{(3)}{}^\mu_{\ \nu}+\cdots,
	\end{eqnarray}
with
\begin{eqnarray}
{\cal K}^{(1)\mu}_{\ \ \ \ \nu} &=&\frac{1}{2}h^{\mu}_{\ \nu}+\Pi^\mu_{\ \nu},\\ 
{\cal K}^{(2)\mu}_{\ \ \ \ \nu} &=&
-\frac{3}{8}h^{\mu\rho}h_{\rho\nu}
-\frac{3}{4}h^{\mu\rho}\Pi_{\rho\nu}
+\frac{1}{4}\Pi^{\mu\rho}h_{\rho\nu}-\frac{1}{2}\bar{\nabla}^\mu \pi^{\rho}\bar{\nabla}_\nu \pi_{\rho}
+\frac{1}{2}\Pi^{\mu\rho}\Pi_{\rho\nu}
+\frac{1}{2}\bar{R}^\mu{}_{\rho\nu\sigma}\pi^\rho\pi^\sigma,\\
{\cal K}^{(3)\mu}_{\ \ \ \ \nu} &=&
\frac{5}{16}h^{\mu\rho}h_{\rho\sigma}h^{\sigma}_{\ \nu}
+\frac{5}{8}h^{\mu\rho}h_{\rho\sigma}\Pi^\sigma_{\ \nu}
-\frac{1}{8}h^{\mu\rho}\Pi_{\rho\sigma}h^{\sigma}_{\ \nu}
-\frac{1}{8}\Pi^{\mu\rho}h_{\rho\sigma}h^{\sigma}_{\ \nu}\notag\\
&&
-\frac{1}{4}h^{\mu\rho}\Pi_{\rho\sigma}\Pi^\sigma_{\ \nu}
-\frac{1}{4}\Pi^{\mu\rho}h_{\rho\sigma}\Pi^\sigma_{\ \nu}
+\frac{1}{4}\Pi^{\mu\rho}\Pi_{\rho\sigma}h^\sigma_{\ \nu}+\frac{3}{8}h^{\mu\rho}\bar{\nabla}_\rho\pi_\sigma \bar{\nabla}_\nu \pi^\sigma
-\frac{1}{8}\bar{\nabla}^\mu\pi_\rho \bar{\nu}^\sigma \pi^\rho h_{\sigma\nu}\notag\\
&&
-\frac{3}{8}h^{\mu\lambda}\bar{R}_{\lambda\rho\nu\sigma}\pi^{\rho}\pi^{\sigma}
+\frac{1}{8}\bar{R}^\mu{}_{\rho\lambda\sigma}\pi^{\rho}\pi^{\sigma}h^{\lambda}_{\ \nu}
\notag\\
&&
+\frac{1}{2}\Pi^{\mu\rho}\Pi_{\rho\sigma}\Pi^{\sigma}_{\ \nu}
-\frac{1}{4}\left(
\Pi^{\mu\rho}\bar{\nabla}_\rho \pi_\sigma \bar{\nabla}_\nu \pi^\sigma
+
\Pi_{\nu}{}^{\rho}\bar{\nabla}_\rho \pi_\sigma \bar{\nabla}^\mu \pi^\sigma
\right)
-\frac{1}{6}\bar{\nabla}_\lambda\bar{R}^\mu{}_{\rho\nu\sigma}\pi^\lambda\pi^\rho\pi^\sigma\notag\\
&&
-\frac{1}{3}\left(
\bar{R}^\mu{}_{\rho\lambda\sigma}\bar{\nabla}_\nu\pi^\lambda
+
\bar{R}_{\nu\rho\lambda\sigma}\bar{\nabla}^\mu\pi^\lambda
\right)
+\frac{1}{4}\left(
\Pi^{\mu\lambda}\bar{R}_{\lambda\rho\nu\sigma}\pi^\rho \pi^\sigma
+
\Pi_{\nu}{}^{\lambda}\bar{R}_{\lambda\rho}{}^\mu{}_{\sigma}\pi^\rho \pi^\sigma
\right)
,
\end{eqnarray}
where $\Pi_{\mu\nu}$ is the symmetric part of $\bar{\nabla}_\mu
\pi_\nu$: 
	\begin{eqnarray}
	\Pi_{\mu\nu} \equiv \frac{1}{2}\left(\bar{\nabla}_{\mu}\pi_{\nu}+\bar{\nabla}_{\nu}\pi_{\mu}\right).
	\end{eqnarray}

Finally, putting everything together and using the expansion for the
determinant, $\sqrt{-g} = \sqrt{-\bar{g}}
\left\{1+h/2+\left(h^2-2h^{\mu\nu}h_{\mu\nu}\right)/8+\cdots\right\}$,
we are able to expand the full dRGT mass terms (\ref{dRGT}) in terms of the metric perturbation $h_{\mu\nu}$ and the Goldstone modes $\pi^{\mu}$ order by order.
At second order, we have  
	\begin{equation}
		\left( \sqrt{-g} 
		{\cal L}^{\mathrm{dRGT}}
		\right) ^{(2)}=\sqrt{-\bar{g}}\left(\mathcal{L}_{h^{2}}+\mathcal{L}_{h\pi}+\mathcal{L}_{\pi^{2}}\right),\label{dRGT_f_2}
	\end{equation}
with
	\begin{eqnarray}
		\mathcal{L}_{h^{2}} & = & \frac{1}{4}\left(h^{2}-h_{\mu\nu}h^{\mu\nu}\right),\label{L2_hh}\\
		\mathcal{L}_{h\pi} & = & h^{\mu\nu}\left(\Pi\bar{g}_{\mu\nu}-\Pi_{\mu\nu}\right),\label{L2_hp}\\
		\mathcal{L}_{\pi^{2}} & =&
		-F_{\mu\nu}F^{\mu\nu}+\bar{R}_{\mu\nu}\pi^{\mu}\pi^{\nu},\label{L2_pp}
	\end{eqnarray}
where $F_{\mu\nu}$ is the antisymmetric part of
$\bar{\nabla}_{\mu}\pi_\nu$,
	\begin{equation}
		F_{\mu\nu} \equiv
		\frac{1}{2}
		\left(
		\bar{\nabla}_{\mu}\pi_{\nu}-\bar{\nabla}_{\nu}\pi_{\mu}
		\right),\label{F_def}
	\end{equation}
and we omitted a total divergence in Eq.~(\ref{L2_pp}).  Similarly, the
Lagrangian at cubic order is
\begin{eqnarray}
\left\{ \sqrt{-g}\mathcal{L}^{\mathrm{dRGT}}\right\} ^{(3)}=\sqrt{-\bar{g}}\left(\mathcal{L}_{h^{3}}+\mathcal{L}_{h^{2}\pi}+\mathcal{L}_{h\pi^{2}}+\mathcal{L}_{\pi^{3}}\right),
\end{eqnarray}
where
\begin{eqnarray}
\mathcal{L}_{h^{3}}&=&\frac{1}{8}\left[\left(1+\alpha_{3}\right)h^{3}-\left(4+3\alpha_{3}\right)hh_{\mu\nu}h^{\mu\nu}+\left(3+2\alpha_{3}\right)h_{\nu}^{\mu}h_{\rho}^{\nu}h_{\mu}^{\rho}\right],\\
\mathcal{L}_{h^{2}\pi}&=&\frac{1}{4}\left[\left(5+6\alpha_{3}\right)h_{\nu}^{\mu}h_{\rho}^{\nu}\Pi_{\mu}^{\rho}-3\left(1+\alpha_{3}\right)h_{\mu\nu}h^{\mu\nu}\Pi-\left(2+3\alpha_{3}\right)h\left(2h^{\nu\rho}\Pi_{\nu\rho}-h\Pi\right)\right],\\
\mathcal{L}_{h\pi^{2}} & = & \frac{1}{2}\left(1+3\alpha_{3}\right)\left[2h^{\mu\nu}\left(\Pi_{\mu}^{\rho}\Pi_{\nu\rho}-\Pi\Pi_{\mu\nu}\right)+h\left(\Pi^{2}-\Pi_{\nu\rho}\Pi^{\nu\rho}\right)\right]\notag\\
 &  & +\frac{1}{2}\left(F_{\mu}^{\rho}h_{\nu\rho}-F_{\mu\nu}h+2\Pi_{\nu\rho}h_{\mu}^{\rho}\right)F^{\mu\nu}+\frac{1}{2}\left(h\bar{R}_{\mu\nu}-h^{\rho\sigma}\bar{R}_{\mu\rho\nu\sigma}\right)\pi^{\mu}\pi^{\nu},
\end{eqnarray}
and
\begin{eqnarray}
\mathcal{L}_{\pi^{3}}=\left(F_{\mu}^{\rho}\Pi_{\nu\rho}-F_{\mu\nu}\Pi\right)F^{\mu\nu}+\left(\bar{R}_{\mu\nu}\Pi-\bar{R}_{\mu\rho\nu\sigma}\Pi^{\rho\sigma}\right)\pi^{\mu}\pi^{\nu}+\alpha_{3}\left(\Pi^{3}-3\Pi\Pi_{\nu\rho}\Pi^{\nu\rho}+2\Pi_{\nu\rho}\Pi_{\mu}^{\rho}\Pi^{\mu\nu}\right).
\end{eqnarray}
At quartic order we have
	\begin{eqnarray}
	\left\{ \sqrt{-g}\mathcal{L}^{\mathrm{dRGT}}\right\} ^{(4)}=\sqrt{-\bar{g}}\left(\mathcal{L}_{h^{4}}+\mathcal{L}_{h^{3}\pi}+\mathcal{L}_{h^{2}\pi^{2}}+\mathcal{L}_{h\pi^{3}}+\mathcal{L}_{\pi^{4}}\right),
	\end{eqnarray}
where
\begin{eqnarray}
\mathcal{L}_{h^{4}} & = & \frac{1}{32}\left(1+2\alpha_{3}+2\alpha_{4}\right)h^{4}-\frac{3}{32}\left(3+5\alpha_{3}+4\alpha_{4}\right)h^{2}h_{\mu\nu}h^{\mu\nu}+\frac{1}{16}\left(8+11\alpha_{3}+8\alpha_{4}\right)hh_{\nu}^{\mu}h_{\rho}^{\nu}h_{\mu}^{\rho}\notag\\
 &  & +\frac{1}{64}\left(13+18\alpha_{3}+12\alpha_{4}\right)\left(h_{\mu\nu}h^{\mu\nu}\right)^{2}-\frac{1}{64}\left(29+36\alpha_{3}+24\alpha_{4}\right)h_{\nu}^{\mu}h_{\rho}^{\nu}h_{\sigma}^{\rho}h_{\mu}^{\sigma},\\
\mathcal{L}_{h^{3}\pi} & = & -\frac{1}{8}\left(11+24\alpha_{3}+24\alpha_{4}\right)h_{\mu}^{\rho}h^{\mu\nu}h_{\nu}^{\sigma}\Pi_{\rho\sigma}+\frac{1}{8}\left(8+21\alpha_{3}+24\alpha_{4}\right)hh_{\nu}^{\sigma}h^{\nu\rho}\Pi_{\rho\sigma}\notag\\
 &  & +\frac{1}{8}\left(5+12\alpha_{3}+12\alpha_{4}\right)h_{\mu\nu}h^{\mu\nu}\left(h^{\rho\sigma}\Pi_{\rho\sigma}-h\Pi\right)+\frac{1}{8}\left(5+9\alpha_{3}+8\alpha_{4}\right)h_{\mu}^{\rho}h^{\mu\nu}h_{\nu\rho}\Pi\notag\\
 &  & +\frac{1}{8}\left(1+3\alpha_{3}+4\alpha_{4}\right)h^{2}\left(h\Pi-3h^{\mu\nu}\Pi_{\mu\nu}\right),\\
\mathcal{L}_{h^{2}\pi^{2}} & = & -\frac{1}{8}\left(5+6\alpha_{3}\right)\left(F_{\lambda}^{\nu}F^{\lambda\mu}h_{\mu}^{\rho}h_{\nu\rho}+2F^{\lambda\mu}h_{\lambda}^{\nu}h_{\nu}^{\rho}\Pi_{\mu\rho}-h_{\nu}^{\sigma}h^{\nu\rho}\bar{R}_{\lambda\rho\mu\sigma}\pi^{\lambda}\pi^{\mu}\right)\notag\\
 &  & +\frac{3}{8}\left(1+\alpha_{3}\right)h_{\nu\rho}h^{\nu\rho}\left(F_{\lambda\mu}F^{\lambda\mu}-\bar{R}_{\lambda\mu}\pi^{\lambda}\pi^{\mu}\right)+\frac{1}{8}\left(2+3\alpha_{3}\right)\left(2F_{\lambda}^{\nu}F^{\lambda\mu}h_{\mu\nu}h-F_{\lambda\mu}F^{\lambda\mu}h^{2}\right)\notag\\
 &  & +\frac{1}{8}\left(2+3\alpha_{3}\right)\left(4F^{\lambda\mu}h_{\lambda}^{\nu}h\Pi_{\mu\nu}+\bar{R}_{\mu\rho\nu\sigma}h\left(h\bar{g}^{\rho\sigma}-2h^{\rho\sigma}\right)\pi^{\mu}\pi^{\nu}\right)\notag\\
 &  & +\frac{1}{8}\left(1+6\alpha_{3}+12\alpha_{4}\right)\left[2\left(h^{\mu\nu}\Pi_{\mu\nu}\right)^{2}+4hh^{\mu\nu}\left(\Pi_{\mu}^{\rho}\Pi_{\nu\rho}-\Pi_{\mu\nu}\Pi\right)+h^{2}\left(\Pi^{2}-\Pi_{\mu\nu}\Pi^{\mu\nu}\right)\right]\notag\\
 &  & +\frac{1}{8}\left(2+9\alpha_{3}+12\alpha_{4}\right)h_{\mu\nu}h^{\mu\nu}\left(\Pi_{\rho\sigma}\Pi^{\rho\sigma}-\Pi^{2}\right)-\frac{1}{8}\left(7+30\alpha_{3}+48\alpha_{4}\right)h_{\lambda}^{\nu}h^{\lambda\mu}\Pi_{\mu}^{\rho}\Pi_{\nu\rho}\notag\\
 &  & +\frac{3}{4}\left(1+5\alpha_{3}+8\alpha_{4}\right)h_{\lambda}^{\nu}h^{\lambda\mu}\Pi_{\mu\nu}\Pi-\frac{1}{8}\left(1+12\alpha_{3}+24\alpha_{4}\right)h^{\lambda\mu}h^{\nu\rho}\Pi_{\lambda\nu}\Pi_{\mu\rho},\\
\mathcal{L}_{h\pi^{3}} & = & \frac{1}{2}\left[3\alpha_{3}F_{\lambda}^{\nu}h+2\left(1+3\alpha_{3}\right)h_{\lambda}^{\nu}\Pi\right]F^{\lambda\mu}\Pi_{\mu\nu}-\frac{1}{2}\left(1+6\alpha_{3}\right)\left(F_{\lambda}^{\nu}F^{\lambda\mu}h_{\mu}^{\rho}\Pi_{\nu\rho}+F^{\lambda\mu}h_{\lambda}^{\nu}\Pi_{\mu}^{\rho}\Pi_{\nu\rho}\right)\notag\\
 &  & +\frac{1}{2}\left(1+3\alpha_{3}\right)\left[F_{\lambda\mu}F^{\lambda\mu}\left(h^{\nu\rho}\Pi_{\nu\rho}-h\Pi\right)+F_{\lambda}^{\nu}F^{\lambda\mu}h_{\mu\nu}\Pi+\left(\bar{g}_{\nu\rho}\bar{R}_{\lambda\mu}\bar{g}_{\alpha\beta}-\bar{R}_{\lambda\mu}\bar{g}_{\nu\alpha}\bar{g}_{\rho\beta}-\bar{R}_{\lambda\nu\mu\rho}\bar{g}_{\alpha\beta}\right)h^{\nu\rho}\Pi^{\alpha\beta}\pi^{\lambda}\pi^{\mu}\right]\notag\\
 &  & +\frac{1}{2}\left(\alpha_{3}+4\alpha_{4}\right)\left[h^{\lambda\mu}\left(3\Pi_{\lambda\mu}\left(\Pi_{\nu\rho}\Pi^{\nu\rho}-\Pi^{2}\right)+6\Pi_{\lambda}^{\nu}\left(\Pi_{\mu\nu}\Pi-\Pi_{\mu}^{\rho}\Pi_{\nu\rho}\right)\right)+h\left(\Pi^{3}-3\Pi\Pi_{\nu\rho}\Pi^{\nu\rho}+2\Pi_{\mu}^{\rho}\Pi^{\mu\nu}\Pi_{\nu\rho}\right)\right]\notag\\
 &  & +\frac{1}{6}\pi^{\lambda}\pi^{\mu}\left[\left(6F^{\nu\rho}h_{\nu}^{\sigma}+\left(7+18\alpha_{3}\right)h^{\nu\rho}\Pi_{\nu}^{\sigma}-\left(4+9\alpha_{3}\right)h\Pi^{\rho\sigma}\right)\bar{R}_{\lambda\rho\mu\sigma}-\left(h\bar{\nabla}_{\nu}\bar{R}_{\lambda\mu}-h^{\rho\sigma}\bar{\nabla}_{\nu}\bar{R}_{\lambda\rho\mu\sigma}\right)\pi^{\nu}\right],
\end{eqnarray}
and
\begin{eqnarray}
\mathcal{L}_{\pi^{4}} & = & \frac{1}{4}F_{\lambda}^{\nu}F^{\lambda\mu}\left[2\left(1-6\alpha_{3}\right)\Pi_{\mu}^{\rho}\Pi_{\nu\rho}-4\left(1-3\alpha_{3}\right)\Pi_{\mu\nu}\Pi-F_{\mu}^{\rho}F_{\nu\rho}\right]+\frac{1}{4}\left(F_{\lambda\mu}F^{\lambda\mu}\right)^{2}+\frac{1}{2}F^{\lambda\mu}F^{\nu\rho}\Pi_{\lambda\nu}\Pi_{\mu\rho}\notag\\
 &  & +\alpha_{4}\left(\Pi^{4}+8\Pi\Pi_{\mu}{}^{\rho}\Pi^{\mu\nu}\Pi_{\nu\rho}+3\left(\Pi_{\lambda\mu}\Pi^{\lambda\mu}\right)^{2}-6\Pi^{2}\Pi_{\nu\rho}\Pi^{\nu\rho}-6\Pi_{\lambda}^{\nu}\Pi^{\lambda\mu}\Pi_{\mu}^{\rho}\Pi_{\nu\rho}\right)\notag\\
 &  & -\frac{3}{2}\alpha_{3}\left(F_{\lambda\mu}F^{\lambda\mu}-\bar{R}_{\lambda\mu}\pi^{\lambda}\pi^{\mu}\right)\left(\Pi^{2}-\Pi_{\nu\rho}\Pi^{\nu\rho}\right)+\frac{1}{6}\bar{R}_{\lambda\rho\mu\sigma}\left(3F_{\nu}^{\sigma}F^{\nu\rho}+2F^{\nu\rho}\Pi_{\nu}^{\sigma}-3F_{\nu\alpha}F^{\nu\alpha}\bar{g}^{\rho\sigma}\right)\pi^{\lambda}\pi^{\mu}\notag\\
 &  & +\frac{1}{3}\left(1+9\alpha_{3}\right)\bar{R}_{\lambda\rho\mu\sigma}\left(\Pi_{\nu}^{\sigma}\Pi^{\nu\rho}-\Pi\Pi^{\rho\sigma}\right)\pi^{\lambda}\pi^{\mu}+\frac{1}{4}\left(\bar{R}_{\lambda\mu}\bar{R}_{\nu\rho}-\bar{R}_{\lambda}{}^{\sigma}{}_{\mu}{}^{\tau}\bar{R}_{\nu\sigma\rho\tau}\right)\pi^{\lambda}\pi^{\mu}\pi^{\nu}\pi^{\rho}\notag\\
 &  & -\frac{1}{3}\bar{\nabla}_{\nu}\left(\bar{g}_{\rho\sigma}\bar{R}_{\lambda\mu}-\bar{R}_{\lambda\rho\mu\sigma}\right)\Pi^{\rho\sigma}\pi^{\lambda}\pi^{\mu}\pi^{\nu}.
\end{eqnarray}

The above expressions are very cumbersome. In the next section, 
we will show that it is possible to introduce a generalized $\Lambda_3$-decoupling limit as in the case of the flat fiducial metric. All terms with cutoff scales lower than $\Lambda_3$ drop out and the resulting terms represent a healthy theory describing various modes propagating on a curved background.

\section{Decoupling limit and the helicity-0 mode} \label{sec:dec}

\subsection{Scales}
The covariant approach employed in the previous sections enables us
to identify the propagating degrees of freedom correctly. We may split
$\pi_{\mu}$ into transverse and longitudinal modes as in the case of the
flat fiducial metric:
	\begin{equation}
		\pi_{\mu} = A_{\mu} + \bar{\nabla}_{\mu}\pi = \frac{1}{M_{\mathrm{pl}} m}\hat{A}_{\mu} +\frac{1}{M_{\mathrm{pl}} m^2}\bar{\nabla}_{\mu} \hat{\pi},
	\end{equation}
where $\hat{A}_{\mu}$ and $\hat{\pi}$ are normalized and are identified
as the helicity-1 and helicity-0 modes, respectively. Similarly, we
define the normalized $\hat{h}_{\mu\nu}$ as
	\begin{equation}
		\hat{h}_{\mu\nu} = M_{\mathrm{pl}} h_{\mu\nu}.
	\end{equation}

The St\"{u}ckelberg expansion yields a whole hierarchy of interaction
terms of $\hat{h}_{\mu\nu}$, $\hat{A}_{\mu}$, and $\hat{\pi}$ with
various energy scales. 
Note that while $h_{\mu\nu}$ without derivatives appear in the
expansion, $\pi_\mu$ without derivatives does not in the case of the
flat fiducial metric.  This point should be contrasted with the curved
case, as $\pi_\mu$ now may appear without derivatives due to the
presence of the curvature tensor and its derivatives, which come
from the commutation of the covariant derivatives. Thus, a general
interaction term takes the following prototype
	\begin{equation}
		M_{\mathrm{pl}}^{2}m^{2}\left(\bar{\nabla}^{d}\bar{R}^{r}\right)h^{n_{h}}A^{a}\left(\bar{\nabla}A\right)^{n_{A}-a}\left(\bar{\nabla}\pi\right)^{2r+d-a}\left(\bar{\nabla}^{2}\pi\right)^{n_{\pi}-2r-d+a}, \label{proto_int}
	\end{equation}
where $n_h$, $n_{A}$, and $n_{\pi}$ are the numbers of the
corresponding fields,
$r$ is the power of curvature terms,
and $d$ is the number of derivatives acting on the curvature.
All the powers in Eq.~(\ref{proto_int}) must be non-negative integers so that, especially,
	\begin{equation}
		0\leq a \leq n_A,\qquad
		0\leq 2r+d-a \leq n_{\pi}.
	\end{equation}
In terms of the normalized variables, Eq.~(\ref{proto_int}) can be written as
	\begin{equation}
		\frac{1}{\Lambda_{\lambda}^{p}}\left(\frac{\bar{\nabla}^{d}\bar{R}^{r}}{m^{2r}}\right)\times\hat{h}^{n_{h}}\hat{A}^{a}\left(\bar{\nabla}\hat{A}\right)^{n_{A}-a}\left(\bar{\nabla}\hat{\pi}\right)^{2r+d-a}\left(\bar{\nabla}^{2}\hat{\pi}\right)^{n_{\pi}+a-2r-d}, \label{proto_nor}
	\end{equation}
where $\Lambda_{\lambda}$ is defined as usual as
$\Lambda_{\lambda}\equiv\left(M_{\mathrm{pl}}m^{\lambda-1}\right)^{1/\lambda}$
with
	\begin{equation}
		p = n_{h}+2n_{A}+3n_{\pi}-4-2r,\qquad \lambda = \frac{n_{h}+2n_{A}+3n_{\pi}-4-2r}{n_{h}+n_{A}+n_{\pi}-2}. \label{lambda_def}
	\end{equation} 
Note that in Eq.~(\ref{proto_nor}) we deliberately separate the
dimensionless (nondynamical) factor $\bar{R}^{r}/m^{2r}$ for later
convenience.  At this point, it is clear that the only difference from
the case of the flat fiducial metric is the presence of the curvature
terms in Eq.~(\ref{proto_nor}), which effectively change the cutoff
scales.  Equations~(\ref{proto_nor}) and~(\ref{lambda_def}) generalize
the expressions for the case of the flat fiducial metric ($r=d=a=0$) to
the curved case.

In Appendix \ref{sub:scales}, we list all possible interaction terms with corresponding cutoff scales, up to fourth order in powers of fields. In general, there are two types
of terms suppressed by scales lower than $\Lambda_{3}$: 
	\begin{equation}
		\frac{1}{\Lambda_{\left(3n_{\pi}-4\right)/\left(n_{\pi}-2\right)}^{3n_{\pi}-4}}\left(\bar{\nabla}^{2}\hat{\pi}\right)^{n_{\pi}},\qquad\frac{1}{\Lambda_{\left(3n_{\pi}-2\right)/\left(n_{\pi}-1\right)}^{3n_{\pi}-2}}\left(\bar{\nabla}\hat{A}\right)\left(\bar{\nabla}^{2}\hat{\pi}\right)^{n_{\pi}},
	\end{equation}
which are exactly the same as the case of the flat fiducial metric.
On the other hand, terms suppressed by $\Lambda_{3}$ are:
	\begin{equation}
		\frac{1}{\Lambda_{3}^{3n_{\pi}-3}}\times\hat{h}\left(\bar{\nabla}^{2}\hat{\pi}\right)^{n_{\pi}},\qquad\frac{1}{\Lambda_{3}^{3n_{\pi}}}\left(\bar{\nabla}\hat{A}\right)^{2}\left(\bar{\nabla}^{2}\hat{\pi}\right)^{n_{\pi}},\qquad\frac{1}{\Lambda_{3}^{3n_{\pi}-6}}\left(\frac{\bar{\nabla}^{d}\bar{R}}{m^{2}}\right)\times\left(\bar{\nabla}\hat{\pi}\right)^{2+d}\left(\bar{\nabla}^{2}\hat{\pi}\right)^{n_{\pi}-2-d},
	\end{equation}
where the last type of terms arises
due to the presence of the curvature of the fiducial metric.
All the other terms are suppressed by scales higher than $\Lambda_{3}$.

\subsection{Extended $\Lambda_3$-decoupling limit}

In the case of the flat fiducial metric, the
``$\Lambda_{\lambda}$-decoupling limit'' is taken as
	\[
		m\rightarrow 0,\qquad M_{\mathrm{pl}}\rightarrow \infty,\qquad \Lambda_{\lambda} \equiv \left(M_{\mathrm{pl}} m^{\lambda -1}\right)^{1/\lambda}= \text{const}.
	\]
For example, the cutoff scale of nonlinear dRGT massive gravity is
$\Lambda_3$. As we shall see, this holds also for the case of a curved
fiducial metric. Thus, we will take an ``extended'' $\Lambda_3$
decoupling limit as
	\begin{equation}
		m\rightarrow 0,\qquad M_{\mathrm{pl}}\rightarrow \infty,\qquad \Lambda_{3} \equiv \left(M_{\mathrm{pl}} m^2\right)^{1/3}= \text{const},\qquad \frac{\bar{R}_{\mu\nu\rho\sigma}}{m^2} \rightarrow \text{finite}.
	\end{equation}
Note that the interactions between $\hat{A}_{\mu}$ and $\hat{\pi}$ start
at cubic order. Therefore, similarly to the analysis of the flat
fiducial metric, we consistently set $\hat{A}_{\mu}=0$ and concentrate
on $\hat{h}_{\mu\nu}$ and $\hat{\pi}$ in the following.

After taking this extended $\Lambda_3$ decoupling limit,
the surviving interaction terms up to fourth order in fields are given by
	\begin{eqnarray}
	\sqrt{-g}M_{\mathrm{pl}}^{2}m^{2}\mathcal{L}^{\text{dRGT}} & \xrightarrow{\text{D.L.}} & \sqrt{-\bar{g}}\bigg[\hat{h}^{\mu\nu}\left(X_{\mu\nu}^{(1)}\left(\hat{\pi}\right)+\frac{1+3\alpha_{3}}{2\Lambda_{3}^{3}}X_{\mu\nu}^{(2)}\left(\hat{\pi}\right)+\frac{\alpha_{3}+4\alpha_{4}}{2\Lambda_{3}^{6}}X_{\mu\nu}^{(3)}\left(\hat{\pi}\right)\right)\nonumber \\
	 &  & \qquad\quad+\mathcal{L}_{\hat{\pi}^{2}}+\mathcal{L}_{\hat{\pi}^{3}}+\mathcal{L}_{\hat{\pi}^{4}}\bigg],\label{L_dl}
	\end{eqnarray}
where 
	\begin{eqnarray}
	X_{\mu\nu}^{(1)}\left(\hat{\pi}\right) & = & \bar{g}_{\mu\nu}\bar{\square}\hat{\pi}-\bar{\nabla}_{\mu}\bar{\nabla}_{\nu}\hat{\pi},\label{X1}\\
	X_{\mu\nu}^{(2)}\left(\hat{\pi}\right) & = & \bar{g}_{\mu\nu}\left(\left(\bar{\square}\hat{\pi}\right)^{2}-\bar{\nabla}_{\rho}\bar{\nabla}_{\sigma}\hat{\pi}\bar{\nabla}^{\rho}\bar{\nabla}^{\sigma}\hat{\pi}\right)+2\left(\bar{\nabla}_{\mu}\bar{\nabla}_{\rho}\hat{\pi}\bar{\nabla}^{\rho}\bar{\nabla}_{\nu}\hat{\pi}-\bar{\square}\hat{\pi}\bar{\nabla}_{\mu}\bar{\nabla}_{\nu}\hat{\pi}\right),\label{X2}\\
	X_{\mu\nu}^{(3)}\left(\hat{\pi}\right) & \equiv & \bar{g}_{\mu\nu}\left(\left(\bar{\square}\hat{\pi}\right)^{3}-3\bar{\square}\hat{\pi}\bar{\nabla}_{\rho}\bar{\nabla}_{\sigma}\hat{\pi}\bar{\nabla}^{\rho}\bar{\nabla}^{\sigma}\hat{\pi}+2\bar{\nabla}^{\rho}\bar{\nabla}_{\sigma}\hat{\pi}\bar{\nabla}^{\sigma}\bar{\nabla}_{\lambda}\hat{\pi}\bar{\nabla}^{\lambda}\bar{\nabla}_{\rho}\hat{\pi}\right)\nonumber \\
	 &  & +3\bar{\nabla}_{\mu}\bar{\nabla}_{\nu}\hat{\pi}\left(\bar{\nabla}_{\rho}\bar{\nabla}_{\sigma}\hat{\pi}\bar{\nabla}^{\rho}\bar{\nabla}^{\sigma}\hat{\pi}-\left(\bar{\square}\hat{\pi}\right)^{2}\right)+6\bar{\nabla}^{\rho}\bar{\nabla}_{\mu}\hat{\pi}\left(\bar{\nabla}_{\nu}\bar{\nabla}_{\rho}\hat{\pi}\bar{\square}\hat{\pi}-\bar{\nabla}_{\nu}\bar{\nabla}^{\sigma}\hat{\pi}\bar{\nabla}_{\rho}\bar{\nabla}_{\sigma}\hat{\pi}\right),\label{X3}
	\end{eqnarray}
and 
	\begin{eqnarray}
	\mathcal{L}_{\hat{\pi}^{2}} & \simeq & \frac{\bar{R}_{\mu\nu}}{m^{2}}\bar{\nabla}^{\mu}\hat{\pi}\bar{\nabla}^{\nu}\hat{\pi},\label{L_dl_pp}\\
	\mathcal{L}_{\hat{\pi}^{3}} & \simeq & \frac{1}{\Lambda_{3}^{3}}\mathcal{A}_{\mu\nu\rho\sigma}\bar{\nabla}^{\mu}\hat{\pi}\bar{\nabla}^{\nu}\hat{\pi}\bar{\nabla}^{\rho}\bar{\nabla}^{\sigma}\hat{\pi},\label{L_dl_ppp}\\
	\mathcal{L}_{\hat{\pi}^{4}} & \simeq & \frac{1}{\Lambda_{3}^{6}}\left(\mathcal{B}_{\mu\nu\rho\sigma\rho'\sigma'}\bar{\nabla}^{\rho'}\bar{\nabla}^{\sigma'}\hat{\pi}-\frac{1}{3}\mathcal{C}_{\lambda\mu\nu\rho\sigma}\bar{\nabla}^{\lambda}\hat{\pi}\right)\bar{\nabla}^{\mu}\hat{\pi}\bar{\nabla}^{\nu}\hat{\pi}\bar{\nabla}^{\rho}\bar{\nabla}^{\sigma}\hat{\pi},\label{L_dl_pppp}
	\end{eqnarray}
with\footnote{Here the (anti)symmetrization is normalized, e.g., $A_{(\mu} B_{\nu)} = \frac{1}{2}\left(A_{\mu}B_{\nu} + A_{\nu}B_{\mu}\right)$ etc.}
	\begin{eqnarray}
	\mathcal{A}_{\mu\nu\rho\sigma} & \equiv & \frac{1}{m^{2}}\left[\left(1+2\alpha_{3}\right)\left(\bar{R}_{\mu\nu}\bar{g}_{\rho\sigma}+\bar{R}_{\rho(\mu\nu)\sigma}\right)-\alpha_{3}\left(\bar{g}_{\rho(\mu}\bar{R}_{\nu)\sigma}+\bar{g}_{\sigma(\mu}\bar{R}_{\nu)\rho}\right)\right],\label{A_def}\\
	\mathcal{B}_{\mu\nu\rho\sigma\rho'\sigma'} & \equiv & \frac{1}{m^{2}}\bigg[\frac{3}{2}\left(\alpha_{3}+2\alpha_{4}\right)\bar{R}_{\mu\nu}\left(2\bar{g}_{\rho[\sigma}\bar{g}_{\sigma']\rho'}\right)+12\alpha_{4}\bar{R}_{\mu[\rho}\bar{g}_{\rho'][\sigma}\bar{g}_{\sigma']\nu}\nonumber \\
	 &  & \qquad\qquad-\frac{1}{3}\left(1+9\alpha_{3}+18\alpha_{4}\right)\left(\bar{R}_{\mu\rho\nu[\sigma}\bar{g}_{\sigma']\rho'}-\bar{R}_{\mu\rho'\nu[\sigma}\bar{g}_{\sigma']\rho}\right)-6\alpha_{4}\bar{g}_{\mu[\rho}\bar{R}_{\rho']\nu\sigma\sigma'}\bigg],\label{B_def}\\
	\mathcal{C}_{\lambda\mu\nu\rho\sigma} & \equiv & \frac{1}{m^{2}}\left[\bar{g}_{\rho\sigma}\bar{\nabla}_{(\lambda}\bar{R}_{\mu\nu)}+\frac{1}{3}\left(\bar{\nabla}_{\lambda}\bar{R}_{\mu(\rho\sigma)\nu}+\bar{\nabla}_{\mu}\bar{R}_{\lambda(\rho\sigma)\nu}+\bar{\nabla}_{\nu}\bar{R}_{\lambda(\rho\sigma)\mu}\right)\right].\label{C_def}
	\end{eqnarray}	
It is not surprising that the self-interactions of $\hat{\pi}$ are all
proportional to the curvature of the fiducial metric, which exactly
vanish in the flat limit.  Note that in deriving Eqs.~(\ref{L_dl_pp}),
(\ref{L_dl_ppp}), and (\ref{L_dl_pppp}), we employed several
integrations by parts, see Appendix \ref{sec:ibp} for details.
For later
convenience, note also that $\mathcal{B}_{\mu\nu\rho\sigma\rho'\sigma'}$ has the
following antisymmetries:
	\begin{equation}
		\mathcal{B}_{\mu\nu\rho\sigma\rho'\sigma'} =  - \mathcal{B}_{\mu\nu\rho'\sigma\rho\sigma'}= - \mathcal{B}_{\mu\nu\rho\sigma'\rho'\sigma}. \label{B_anti}
	\end{equation}

\subsection{Unmixing $\hat{h}_{\mu\nu}$ and $\hat{\pi}$}

Unlike the case of the flat fiducial metric, here $\hat{\pi}$ acquires a
quadratic kinetic term (\ref{L_dl_pp}) automatically, due to the
nonvanishing curvature of the fiducial metric. Nevertheless, it is
interesting to perform the field redefinition
	\begin{equation}
		\hat{h}_{\mu\nu} \rightarrow \hat{h}_{\mu\nu} + \hat{\pi} \bar{g}_{\mu\nu} - \frac{1+3\alpha_3}{\Lambda_3^3} \bar{\nabla}_{\mu}\hat{\pi}\bar{\nabla}_{\nu}\hat{\pi},
	\end{equation}
under which $\hat{h}^{\mu\nu}X_{\mu\nu}^{(1)}$ and
$\hat{h}^{\mu\nu}X_{\mu\nu}^{(2)}$ get unmixed as in the case of the flat
fiducial metric. To this end, we first expand the Einstein-Hilbert
Lagrangian around a general background up to quadratic order,
	\begin{equation}
		\left\{ \frac{1}{2}M_{\mathrm{pl}}^{2}\sqrt{-g}R\right\} _{2}\simeq-\frac{1}{4}\sqrt{-\bar{g}}\hat{h}_{\mu\nu}\mathcal{E}^{\mu\nu,\rho\sigma}\hat{h}_{\rho\sigma},
	\end{equation}
where the ``Lichnerowicz operator'' is defined by
	\begin{eqnarray}
	\mathcal{E}_{\mu\nu,\rho\sigma}\hat{h}^{\rho\sigma} & \equiv & -\frac{1}{2}\bar{\square}\hat{h}_{\mu\nu}-\frac{1}{2}\bar{\nabla}_{\mu}\bar{\nabla}_{\nu}\hat{h}+\frac{1}{2}\bar{g}_{\mu\nu}\left(\bar{\square}\hat{h}-\bar{\nabla}_{\rho}\bar{\nabla}_{\sigma}\hat{h}^{\rho\sigma}\right)+\bar{\nabla}_{\rho}\bar{\nabla}_{(\mu}\hat{h}_{\nu)}^{\rho}\nonumber \\
	 &  & -2\left(\hat{h}_{(\mu}^{\rho}\bar{R}_{\nu)\rho}-\frac{1}{2}\hat{h}\bar{R}_{\mu\nu}\right)-\frac{1}{4}\left(\bar{g}_{\mu\nu}\hat{h}-2\hat{h}_{\mu\nu}\right)\bar{R}.\label{L_EH_linear}
	\end{eqnarray}
After taking the $\Lambda_3$ decoupling limit, the terms in the second
line of Eq.~(\ref{L_EH_linear}) drop out and thus would not contribute to
the (partially) unmixed Lagrangian.

With some manipulations, the final Lagrangian can be written as
	\begin{equation}
		\mathcal{L}=\mathcal{L}_{2}+\mathcal{L}_{3}+\mathcal{L}_{4},
	\end{equation}
where the quadratic terms are
	\begin{equation}
		\mathcal{L}_{2}=-\frac{1}{4}\hat{h}_{\mu\nu}\mathcal{E}^{\mu\nu,\rho\sigma}\hat{h}_{\rho\sigma}-\frac{1}{2}\left(\frac{3}{2}\bar{g}_{\mu\nu}-\frac{\bar{R}_{\mu\nu}}{m^{2}}\right)\bar{\nabla}^{\mu}\hat{\pi}\bar{\nabla}^{\nu}\hat{\pi}. \label{L_unmix_2}
	\end{equation}
It is interesting to see that, at linear order, $\hat{\pi}$ propagates
in an effective metric $(3/2)\bar{g}_{\mu\nu}-
\bar{R}_{\mu\nu}/m^{2}$. Thus, although there are no higher derivatives
so that the theory is free of any extra modes, $\hat{\pi}$ itself is a
ghost in spacetime regions where
	\begin{equation}
		\bar{R}_{\mu\nu} > \frac{3}{2} m^2\bar{g}_{\mu\nu}, \label{gfcon}
	\end{equation}
	is satisfied.
This generalizes the well-known ``Higuchi bound'' in the de Sitter
background~\cite{Higuchi:1986py}. A critical case arises for
$\bar{R}_{\mu\nu} = (3/2) m^2\bar{g}_{\mu\nu}$, where $\hat{\pi}$
becomes nondynamical (at the linear level). This case corresponds
to the case of ``partially massless'' gravity.

The cubic and quartic parts are
	\begin{equation}
	\mathcal{L}_{3}=-\frac{3\left(1+3\alpha_{3}\right)}{4\Lambda_{3}^{3}}\left(\bar{\nabla}\hat{\pi}\right)^{2}\bar{\square}\hat{\pi}+\frac{1}{2\Lambda_{3}^{3}}\mathcal{A}_{\mu\nu\rho\sigma}\bar{\nabla}^{\mu}\hat{\pi}\bar{\nabla}^{\nu}\hat{\pi}\bar{\nabla}^{\rho}\bar{\nabla}^{\sigma}\hat{\pi},\label{L_unmix_3}
	\end{equation}
and
	\begin{eqnarray}
	\mathcal{L}_{4} & = & -\frac{1+8\alpha_{3}+9\alpha_{3}^{2}+8\alpha_{4}}{4\Lambda_{3}^{6}}\left(\bar{\nabla}\hat{\pi}\right)^{2}\left(\left(\bar{\square}\hat{\pi}\right)^{2}-\bar{\nabla}_{\rho}\bar{\nabla}_{\sigma}\hat{\pi}\bar{\nabla}^{\rho}\bar{\nabla}^{\sigma}\hat{\pi}\right)+\frac{1}{4\Lambda_{3}^{6}}\left(\alpha_{3}+4\alpha_{4}\right)\hat{h}^{\mu\nu}X_{\mu\nu}^{(3)}\left(\hat{\pi}\right)\nonumber \\
	 &  & +\frac{1}{2\Lambda_{3}^{6}}\left(\mathcal{B}_{\mu\nu\rho\sigma\rho'\sigma'}\bar{\nabla}^{\rho'}\bar{\nabla}^{\sigma'}\hat{\pi}-\frac{1}{3}\mathcal{C}_{\lambda\mu\nu\rho\sigma}\bar{\nabla}^{\lambda}\hat{\pi}\right)\bar{\nabla}^{\mu}\hat{\pi}\bar{\nabla}^{\nu}\hat{\pi}\bar{\nabla}^{\rho}\bar{\nabla}^{\sigma}\hat{\pi},\label{L_unmix_4}
	\end{eqnarray}	
respectively, where $\mathcal{A}_{\mu\nu\rho\sigma}$ etc.
are defined in Eqs.~(\ref{A_def})--(\ref{C_def}).

In the case of the flat fiducial metric, a necessary condition for the
absence of the BD ghost is the disappearance of self-interactions of the
helicity-0 mode $\hat{\pi}$. This is because there $\hat{\pi}$ appears
always with two derivatives, $\partial_{\mu}\partial_{\nu} \hat{\pi}$,
and thus any self-interaction of $\hat{\pi}$ will inevitably yield
higher derivatives in the equations of motion. In our case, $\hat{\pi}$ acquires
self-interactions due to the presence of the curvature tensor of the
fiducial metric.  However, in the extended $\Lambda_3$ decoupling limit,
the equation of motion for $\hat{\pi}$ remains of second order in
derivatives (acting on $\hat{\pi}$).  To see this, first note that, for
$\mathcal{L}_3$ and for the term proportional to
$\mathcal{C}_{\lambda\mu\nu\rho\sigma}$ in $\mathcal{L}_4$, the second
derivatives of $\hat{\pi}$ appear linearly, implying that the
corresponding equation of motion for $\hat{\pi}$ is of second order in
derivatives.  As for the first term in $\mathcal{L}_4$, though it does
not take the form of covariant Galileons, which must be supplemented
with a curvature term $\sim \bar{R}
\left(\bar{\nabla}\hat{\pi}\right)^4$, it is straightforward to check
that the corresponding equation of motion for $\hat{\pi}$ is of second
order. The point is that the equation of  motion contains derivatives of the
curvature of the fiducial metric, which are definitely safe since the
fiducial metric is nondynamical. This is the same for the terms
proportional to $\hat{h}^{\mu\nu}X^{(3)}_{\mu\nu}(\hat{\pi})$ and
$\mathcal{B}_{\mu\nu\rho\sigma\rho'\sigma'}$ due to the antisymmetries
(\ref{B_anti}), see Appendix \ref{sec:eom} for explicit proofs.  To
summarize, similarly to the case of the flat fiducial metric, in the
$\Lambda_3$ decoupling limit, $\hat{\pi}$ propagates subject to a
second-order equation of motion, which prevents the BD ghost.

\section{Conclusion}

In this paper, we have extended the St\"{u}ckelberg analysis, which was used
in the theory on the flat fiducial metric, to the theory on a general
fiducial metric. First, we have given the covariant definition of the
perturbation $\pi^\mu$ of the St\"{u}ckelberg field. Using this
definition, we have expanded the action in a covariant way and given the
explicit expression for the action of $h_{\mu\nu}$ and $\pi^\mu$ up to
fourth order.

As an application of this formula, we have calculated the action of the
helicity-0 mode $\pi$. From the second-order action, we have obtained
the ghost-free condition~(\ref{gfcon}), which is the generalization of
the Higuchi bound known in the de Sitter fiducial case.  Contrary to the
flat fiducial case, we have faced the problem in taking the $\Lambda_3$
decoupling limit in the general fiducial case. However, we have overcome
this problem by extending the $\Lambda_3$ decoupling limit, in which the
curvature of the fiducial metric is scaled. In this extended $\Lambda_3$
decoupling limit, the helicity-0 mode $\pi$ and helicity-2 mode
$h_{\mu\nu}$ are decoupled as in the flat fiducial case. (Of course,
there remains the $h^{\mu\nu}X_{\mu\nu}^{(3)}$ coupling term, which
exists even in the flat fiducial case.) The decoupled action is composed
of the flat result and the curvature correction. The most important
result is that this curvature correction does not produce any higher
derivatives, leading to the second-order equation of motion.  This fact
offers us a different and complementary way of clarifying the reason for
the absence of the BD ghost in dRGT massive gravity on a general
fiducial metric. If we go into bigravity, the fiducial metric
becomes dynamical as well. Then, we need to take into account the
equation of motion for the fiducial (dynamical) metric. This is still an
open question and left for further investigation.

\acknowledgments

We would like to thank K. Hinterbichler and S. Renaux-Petel for
discussion, C. de Rham for correspondence and S. F. Hassan for useful comments.
 This work was in part
supported by the JSPS Grant-in-Aid for Scientific Research No. 24740161
(T.K.), No. 25287054 (X.G. and M.Y.), and No. 26610062 (M.Y.), and the JSPS
Research Fellowship for Young Scientists, No. 26-11495 (D.Y.).

\appendix

\section{ANOTHER APPROACH TO EXPAND FIDUCIAL METRIC}
\label{sc:fieldspace}

In the main text, we regard the St\"{u}ckelberg fields as
diffeomorphisms of the physical spacetime. From an alternative point of
view, however, $\phi^a$ are simply four scalar fields living on the
physical spacetime, which form a four-dimensional field space at each
spacetime point.  It is thus interesting to see whether the relation
(\ref{f_exp_lie}) can be reproduced from this point of view.

To this end, first we introduce the metric $\bar{g}_{ab}$ in the field space as
	\begin{eqnarray}
		f_{\mu\nu}(\phi) = \bar{g}_{ab}(\phi)\frac{\partial \phi^a(x)}{\partial x^\mu}\frac{\partial \phi^b(x)}{\partial x^\mu}. \label{ffs}
	\end{eqnarray} 
The one parameter family of map $\phi_\lambda$ defined in
Sec.\ref{CovDefpi} corresponds to curves in the field space, which can
be written as $\phi^a(\lambda)$.  The unitary gauge is chosen such that
$\phi^a(0)= x^\mu \delta^a_\mu$.  Expanding (\ref{ffs}) around the
unitary gauge and then setting $\lambda=-1$, we have
\begin{eqnarray}
f_{\mu\nu} \equiv \left.f_{\mu\nu}(\phi)\right|_{\lambda=-1}
=\left.\mathrm{e}^{-\frac{\mathrm{d}}{\mathrm{d}\lambda}}f_{\mu\nu}(\phi(\lambda))\right|_{\lambda=0}.
\end{eqnarray}
Since $f_{\mu\nu}$ is a scalar in field space,
$\mathrm{d}/\mathrm{d}\lambda$ can be replaced by the Lie derivative along
the curve:
\begin{eqnarray}
f_{\mu\nu}(\phi) =\left. \mathrm{e}^{\mathsterling_u}f_{\mu\nu}\right|_{\lambda=0},\label{fisexpliedel}
\end{eqnarray}
where $u^a$ is the tangent vector of the curve $\phi^a(\lambda)$.
Moreover, (\ref{deltaamu}) implies
\begin{eqnarray}
\mathsterling_{u} 
\left(
\frac{\partial \phi^a(\lambda)}{\partial x^\mu}
\right) = 0.
\end{eqnarray}
Thus in (\ref{fisexpliedel}), the Lie derivative acts only on $\bar{g}_{ab}$, i.e.
\begin{eqnarray}
f_{\mu\nu}
=
\left. 
\left(
\left(
\mathrm{e}^{\mathsterling_{u}}\bar{g}_{ab}
\right)
\frac{\partial \phi^a(\lambda)}{\partial x^\mu}
\frac{\partial \phi^b(\lambda)}{\partial x^\nu}
\right)
\right|_{\lambda'=0}
=
\left.
\left(
\mathrm{e}^{\mathsterling_{u}}\bar{g}_{ab}
\right)
\right|_{\lambda=0} \delta^a_\mu\delta^b_\nu,
\end{eqnarray}
which exactly coincides with (\ref{f_exp_lie}).

\section{INTERACTION TERMS AND CUTOFF SCALES}\label{sub:scales}

For completeness, according to (\ref{proto_nor}), here we list all
possible interaction terms in the St\"{u}ckelberg expansion up to the
fourth order in fields as well as their corresponding cutoff scales.

\begin{itemize}
\item $n_{h}+n_{A}+n_{\pi}=2$

\begin{center}
\begin{tabular}{c|c|c|c|c|c|c|c|c|c}
\hline 
$\left(n_{h},n_{A},n_{\pi}\right)$ & $\left(2,0,0\right)$ & \multicolumn{2}{c|}{$\left(0,2,0\right)$} & \multicolumn{2}{c|}{$\left(0,0,2\right)$} & $\left(1,1,0\right)$ & $\left(1,0,1\right)$ & \multicolumn{2}{c}{$\left(0,1,1\right)$}\tabularnewline
\hline 
$\left(r,d,a\right)$ & $\left(0,0,0\right)$ & $\left(0,0,0\right)$ & $\left(1,0,2\right)$ & $\left(0,0,0\right)$ & $\left(1,0,0\right)$ & $\left(0,0,0\right)$ & $\left(0,0,0\right)$ & $\left(0,0,0\right)$ & $\left(1,0,1\right)$\tabularnewline
\hline 
$\Lambda_{\lambda}^{p}$ & $m^{-2}$ & 1 & $m^{-2}$ & $m^{2}$ & 1 & $m^{-1}$ & 1 & $m$ & $m^{-1}$\tabularnewline
\hline 
\end{tabular}
\par\end{center}

In this case, there are two types of terms with lowest scales:
	\[
		\frac{1}{m^2} \left(\bar{\nabla}^2 \hat{\pi}\right)^2,\qquad \frac{1}{m}\left(\bar{\nabla}\hat{A}\right)\left(\bar{\nabla}^2 \hat{\pi}\right),
	\]
while there are three types of terms
	\[
		\hat{h}\left(\bar{\nabla}^2\hat{\pi}\right),\qquad
		\left(\bar{\nabla}\hat{A}\right)^2,\qquad \frac{\bar{R}}{m^2}\left(\bar{\nabla}\hat{\pi}\right)^2, 
	\]
that are ``scale invariant''. 

\item $n_{h}+n_{A}+n_{\pi}=3$

\begin{center}
\begin{tabular}{c|c|c|c|c|c|c|c|cc}
\cline{1-9} 
$\left(n_{h},n_{A},n_{\pi}\right)$ & $\left(3,0,0\right)$ & \multicolumn{3}{c|}{$\left(0,3,0\right)$} & \multicolumn{3}{c|}{$\left(0,0,3\right)$} & $\left(2,1,0\right)$ & \tabularnewline
\cline{1-9} 
$\left(r,d,a\right)$ & $\left(0,0,0\right)$ & $\left(0,0,0\right)$ & $\left(1,0,2\right)$ & $\left(1,1,3\right)$ & $\left(0,0,0\right)$ & $\left(1,0,0\right)$ & $\left(1,1,0\right)$ & $\left(0,0,0\right)$ & \tabularnewline
\cline{1-9} 
$\Lambda_{\lambda}^{p}$ & $M_{\mathrm{pl}}/m^{2}$ & $\Lambda_{2}^{2}$ & \multicolumn{2}{c|}{$M_{\mathrm{pl}}/m$} & $\Lambda_{5}^{5}$ & \multicolumn{2}{c|}{$\Lambda_{3}^{3}$} & $M_{\mathrm{pl}}/m$& \tabularnewline
\hline 
\hline 
$\left(n_{h},n_{A},n_{\pi}\right)$ & \multicolumn{2}{c|}{$\left(1,2,0\right)$} & $\left(2,0,1\right)$ & \multicolumn{2}{c|}{$\left(1,0,2\right)$} & \multicolumn{4}{c}{$\left(0,2,1\right)$}\tabularnewline
\hline 
$\left(r,d,a\right)$ & $\left(0,0,0\right)$ & $\left(1,0,2\right)$ & $\left(0,0,0\right)$ & $\left(0,0,0\right)$ & $\left(1,0,0\right)$ & $\left(0,0,0\right)$ & $\left(1,0,1\right)$ & \multicolumn{1}{c|}{$\left(1,0,2\right)$} & $\left(1,1,2\right)$\tabularnewline
\hline 
$\Lambda_{\lambda}^{p}$ & $M_{\mathrm{pl}}$ & $M_{\mathrm{pl}}/m^{2}$ & $M_{\mathrm{pl}}$ & $\Lambda_{3}^{3}$ & $M_{\mathrm{pl}}$ & $\Lambda_{3}^{3}$ & \multicolumn{3}{c}{$M_{\mathrm{pl}}$}\tabularnewline
\hline 
\hline 
$\left(n_{h},n_{A},n_{\pi}\right)$ & \multicolumn{4}{c|}{$\left(0,1,2\right)$} & \multicolumn{2}{c}{$\left(1,1,1\right)$} & \multicolumn{1}{c}{} &  & \tabularnewline
\cline{1-7} 
$\left(r,d,a\right)$ & $\left(0,0,0\right)$ & $\left(1,0,0\right)$ & $\left(1,0,1\right)$ & $\left(1,1,1\right)$ & $\left(0,0,0\right)$ & \multicolumn{1}{c}{$\left(1,0,1\right)$} & \multicolumn{1}{c}{} &  & \tabularnewline
\cline{1-7} 
$\Lambda_{\lambda}^{p}$ & $\Lambda_{4}^{4}$ & \multicolumn{3}{c|}{$\Lambda_{2}^{2}$} & $\Lambda_{2}^{2}$ & \multicolumn{1}{c}{$M_{\mathrm{pl}}/m$} & \multicolumn{1}{c}{} &  & \tabularnewline
\cline{1-7} 
\end{tabular}
\par\end{center}

The most relevant terms are
	\[
		\frac{1}{\Lambda_5^5}\left(\bar{\nabla}^2\hat{\pi}\right)^3,\qquad 
		\frac{1}{\Lambda_4^4} \left(\bar{\nabla}\hat{A}\right) \left(\bar{\nabla}^2\hat{\pi}\right)^2,
	\]
which have cutoff scales lower than $\Lambda_3$. Terms suppressed by $\Lambda_3$ are
	\[
		\frac{1}{\Lambda_3^3} \hat{h}\left(\bar{\nabla}^2 \hat{\pi}\right)^2,\qquad
		\frac{1}{\Lambda_3^3} \left(\bar{\nabla} \hat{A}\right)^2 \bar{\nabla}^2\hat{\pi},\qquad
		\frac{1}{\Lambda_3^3} \frac{\bar{R}}{m^2} \bar{\nabla}^2\hat{\pi}\left(\bar{\nabla}\hat{\pi}\right)^2,\qquad 
				\frac{1}{\Lambda_3^3} \frac{\bar{\nabla}\bar{R}}{m^2} \left(\bar{\nabla}\hat{\pi}\right)^3,
	\]
where the last two types of terms arise due to the presence of curvature of the fiducial metric.

\item $n_{h}+n_{A}+n_{\pi}=4$

\begin{center}
\begin{tabular}{c|c|c|c|c|c|c|c|cccc}
\hline 
$\left(n_{h},n_{A},n_{\pi}\right)$ & $\left(4,0,0\right)$ & \multicolumn{5}{c|}{$\left(0,4,0\right)$} & \multicolumn{5}{c}{$\left(0,0,4\right)$}\tabularnewline
\hline 
$\left(r,d,a\right)$ & $\left(0,0,0\right)$ & $\left(0,0,0\right)$ & $\left(1,0,2\right)$ & $\left(1,1,3\right)$ & $\left(1,2,4\right)$ & $\left(2,0,4\right)$ & $\left(0,0,0\right)$ & \multicolumn{1}{c|}{$\left(1,0,0\right)$} & \multicolumn{1}{c|}{$\left(1,1,0\right)$} & \multicolumn{1}{c|}{$\left(1,2,0\right)$} & $\left(2,0,0\right)$\tabularnewline
\hline 
$\Lambda_{\lambda}^{p}$ & $M_{\mathrm{pl}}^2/m^{2}$ & $\Lambda_{2}^{4}$ & \multicolumn{3}{c|}{$M_{\mathrm{pl}}^{2}$} & $M_{\mathrm{pl}}^2/m^{2}$ & $\Lambda_{4}^{8}$ & \multicolumn{3}{c|}{$\Lambda_{3}^{6}$} & $\Lambda_{2}^{4}$\tabularnewline
\hline 
\hline 
$\left(n_{h},n_{A},n_{\pi}\right)$ & $\left(3,1,0\right)$ & \multicolumn{3}{c|}{$\left(1,3,0\right)$} & $\left(3,0,1\right)$ & \multicolumn{3}{c}{$\left(1,0,3\right)$} &  &  & \tabularnewline
\cline{1-9} 
$\left(r,d,a\right)$ & $\left(0,0,0\right)$ & $\left(0,0,0\right)$ & $\left(1,0,2\right)$ & $\left(1,1,3\right)$ & $\left(0,0,0\right)$ & $\left(0,0,0\right)$ & $\left(1,0,0\right)$ & $\left(1,1,0\right)$ &  &  & \tabularnewline
\cline{1-9} 
$\Lambda_{\lambda}^{p}$ & $\Lambda_{1/2}$ & $\Lambda_{3/2}^{3}$ & \multicolumn{2}{c|}{$\Lambda_{1/2}$} & $M_{\mathrm{pl}}^{2}$ & $\Lambda_{3}^{6}$ & \multicolumn{2}{c}{$\Lambda_{2}^{4}$} &  &  & \tabularnewline
\hline 
\hline 
$\left(n_{h},n_{A},n_{\pi}\right)$ & \multicolumn{7}{c}{$\left(0,3,1\right)$} &  &  &  & \tabularnewline
\cline{1-8} 
$\left(r,d,a\right)$ & $\left(0,0,0\right)$ & $\left(1,0,1\right)$ & $\left(1,0,2\right)$ & $\left(1,1,2\right)$ & $\left(1,1,3\right)$ & $\left(1,2,3\right)$ & \multicolumn{1}{c}{$\left(2,0,3\right)$} &  &  &  & \tabularnewline
\cline{1-8} 
$\Lambda_{\lambda}^{p}$ & $\Lambda_{5/2}^{5}$ & \multicolumn{5}{c|}{$\Lambda_{3/2}^{3}$} & \multicolumn{1}{c}{$\Lambda_{1/2}$} &  &  &  & \tabularnewline
\hline 
\hline 
$\left(n_{h},n_{A},n_{\pi}\right)$ & \multicolumn{7}{c|}{$\left(0,1,3\right)$} & \multicolumn{2}{c|}{$\left(2,2,0\right)$} & \multicolumn{2}{c}{$\left(2,0,2\right)$}\tabularnewline
\hline 
$\left(r,d,a\right)$ & $\left(0,0,0\right)$ & $\left(1,0,0\right)$ & $\left(1,0,1\right)$ & $\left(1,1,0\right)$ & $\left(1,1,1\right)$ & $\left(1,2,1\right)$ & $\left(2,0,1\right)$ & \multicolumn{1}{c|}{$\left(0,0,0\right)$} & \multicolumn{1}{c|}{$\left(1,0,2\right)$} & \multicolumn{1}{c|}{$\left(0,0,0\right)$} & $\left(1,0,0\right)$\tabularnewline
\hline 
$\Lambda_{\lambda}^{p}$ & $\Lambda_{7/2}^{7}$ & \multicolumn{5}{c|}{$\Lambda_{5/2}^{5}$} & $\Lambda_{3/2}^{3}$ & \multicolumn{1}{c|}{$M_{\mathrm{pl}}^{2}$} & \multicolumn{1}{c|}{$M_{\mathrm{pl}}^2/m^{2}$} & \multicolumn{1}{c|}{$\Lambda_{2}^{4}$} & $M_{\mathrm{pl}}^{2}$\tabularnewline
\hline 
\hline 
$\left(n_{h},n_{A},n_{\pi}\right)$ & \multicolumn{8}{c|}{$\left(0,2,2\right)$} & \multicolumn{2}{c}{$\left(2,1,1\right)$} & \tabularnewline
\cline{1-11} 
$\left(r,d,a\right)$ & $\left(0,0,0\right)$ & $\left(1,0,0\right)$ & $\left(1,0,1\right)$ & $\left(1,0,2\right)$ & $\left(1,1,1\right)$ & $\left(1,1,2\right)$ & $\left(1,2,2\right)$ & \multicolumn{1}{c|}{$\left(2,0,2\right)$} & \multicolumn{1}{c|}{$\left(0,0,0\right)$} & $\left(1,0,1\right)$ & \tabularnewline
\cline{1-11} 
$\Lambda_{\lambda}^{p}$ & $\Lambda_{3}^{6}$ & \multicolumn{6}{c|}{$\Lambda_{2}^{4}$} & \multicolumn{1}{c|}{$M_{\mathrm{pl}}^{2}$} & \multicolumn{1}{c|}{$\Lambda_{3/2}^{3}$} & $\Lambda_{1/2}$ & \tabularnewline
\hline 
\hline 
$\left(n_{h},n_{A},n_{\pi}\right)$ & \multicolumn{4}{c|}{$\left(1,2,1\right)$} & \multicolumn{4}{c}{$\left(1,1,2\right)$} &  &  & \tabularnewline
\cline{1-9} 
$\left(r,d,a\right)$ & $\left(0,0,0\right)$ & $\left(1,0,1\right)$ & $\left(1,0,2\right)$ & $\left(1,1,2\right)$ & $\left(0,0,0\right)$ & $\left(1,0,0\right)$ & $\left(1,0,1\right)$ & $\left(1,1,1\right)$ &  &  & \tabularnewline
\cline{1-9} 
$\Lambda_{\lambda}^{p}$ & $\Lambda_{2}^{4}$ & \multicolumn{3}{c|}{$M_{\mathrm{pl}}^{2}$} & $\Lambda_{5/2}^{5}$ & \multicolumn{3}{c}{$\Lambda_{3/2}^{3}$} &  &  & \tabularnewline
\cline{1-9} 
\end{tabular}
\par\end{center}
In this case, the most relevant terms are
	\[
		\frac{1}{\Lambda_4^8} \left(\bar{\nabla}^4\hat{\pi}\right)^4,\qquad 
		\frac{1}{\Lambda_{7/2}^7} \left(\bar{\nabla}\hat{A}\right) \left(\bar{\nabla}^4\hat{\pi}\right)^3,
	\]
while terms suppressed by $\Lambda_3$ are
	\begin{eqnarray*}
	 &  & \frac{1}{\Lambda_{3}^{6}}\hat{h}\left(\bar{\nabla}^{2}\hat{\pi}\right)^{3},\qquad\frac{1}{\Lambda_{3}^{6}}\left(\bar{\nabla}\hat{A}\right)^{2}\left(\bar{\nabla}^{2}\hat{\pi}\right)^{2},\\
	 &  & \frac{1}{\Lambda_{3}^{6}}\frac{\bar{R}}{m^{2}}\left(\bar{\nabla}\hat{\pi}\right)^{2}\left(\bar{\nabla}^{2}\hat{\pi}\right)^{2},\qquad\frac{1}{\Lambda_{3}^{6}}\frac{\bar{\nabla}\bar{R}}{m^{2}}\left(\bar{\nabla}\hat{\pi}\right)^{3}\left(\bar{\nabla}^{2}\hat{\pi}\right),\qquad\frac{1}{\Lambda_{3}^{6}}\frac{\bar{\nabla}^{2}\bar{R}}{m^{2}}\left(\bar{\nabla}\hat{\pi}\right)^{4}.
	\end{eqnarray*}

\end{itemize}

\section{EQUATION OF MOTION FOR $\hat{\pi}$ IN THE DECOUPLING LIMIT} \label{sec:eom}

First we will show that, as long as the second derivatives of $\hat{\pi}$ enter the Lagrangian linearly, the corresponding equations of motion for $\hat{\pi}$ are up to second order in derivatives.
To this end, consider a general Lagrangian
	\begin{equation}
		\bar{T}_{\lambda_{1}\cdots\lambda_{n}\mu\nu}\bar{\nabla}^{\lambda_{1}}\hat{\pi}\cdots\bar{\nabla}^{\lambda_{n}}\hat{\pi}\bar{\nabla}^{\mu}\bar{\nabla}^{\nu}\hat{\pi},
	\end{equation}
where $\bar{T}_{\lambda_1\cdots \lambda_n \mu\nu}$ contains no
$\hat{\pi}$.  Simple manipulation yields the equation of motion, which
reads
	\begin{eqnarray}
	0 & = & -\sum_{i=1}^{n}\bar{\nabla}^{\lambda_{i}}\left(T_{\lambda_{1}\cdots\lambda_{i}\cdots\lambda_{n}\mu\nu}\bar{\nabla}^{\lambda_{1}}\hat{\pi}\cdots\bar{\nabla}^{\lambda_{i}-1}\hat{\pi}\bar{\nabla}^{\lambda_{i}+1}\hat{\pi}\cdots\bar{\nabla}^{\lambda_{n}}\hat{\pi}\right)\bar{\nabla}^{\mu}\bar{\nabla}^{\nu}\hat{\pi}+\bar{\nabla}^{\mu}\left(\bar{\nabla}^{\nu}T_{\lambda_{1}\cdots\lambda_{n}\mu\nu}\bar{\nabla}^{\lambda_{1}}\hat{\pi}\cdots\bar{\nabla}^{\lambda_{n}}\hat{\pi}\right)\nonumber \\
	 &  & +\sum_{i=1}^{n}\bar{\nabla}^{\mu}\left(T_{\lambda_{1}\cdots\lambda_{i}\cdots\lambda_{n}\mu\nu}\bar{\nabla}^{\lambda_{1}}\hat{\pi}\cdots\bar{\nabla}^{\lambda_{i}-1}\hat{\pi}\bar{\nabla}^{\lambda_{i}+1}\hat{\pi}\cdots\bar{\nabla}^{\lambda_{n}}\hat{\pi}\right)\bar{\nabla}^{\nu}\bar{\nabla}^{\lambda_{i}}\hat{\pi}\nonumber \\
	 &  & +\sum_{i=1}^{n}T_{\lambda_{1}\cdots\lambda_{i}\cdots\lambda_{n}\mu\nu}\bar{\nabla}^{\lambda_{1}}\hat{\pi}\cdots\bar{\nabla}^{\lambda_{i}-1}\hat{\pi}\bar{\nabla}^{\lambda_{i}+1}\hat{\pi}\cdots\bar{\nabla}^{\lambda_{n}}\hat{\pi}\bar{R}^{\mu\lambda_{i}\nu\rho}\bar{\nabla}_{\rho}\hat{\pi},\label{eom1}
	\end{eqnarray}
and hence contains no higher derivatives of $\hat{\pi}$.

Next, for arbitrary tensors $E_{\mu\nu\rho_1\sigma_1\rho_2\sigma_2}$ and $E_{\mu\nu\rho_1\sigma_1\rho_2\sigma_2\rho_3\sigma_3}$ containing no $\hat{\pi}$, which are antisymmetric under exchange of $\rho_i \leftrightarrow \rho_j$ and $\sigma_i \leftrightarrow \sigma_j$ and symmetric under exchange of pairs of $(\rho_i\sigma_i)\leftrightarrow (\rho_j\sigma_j)$, the corresponding equation of motion for $\hat{\pi}$ of
	\begin{equation}
		E_{\mu\nu\rho_{1}\sigma_{1}\rho_{2}\sigma_{2}}\bar{\nabla}^{\mu}\hat{\pi}\bar{\nabla}^{\nu}\hat{\pi}\bar{\nabla}^{\rho_{1}}\bar{\nabla}^{\sigma_{1}}\hat{\pi}\bar{\nabla}^{\rho_{2}}\bar{\nabla}^{\sigma_{2}}\hat{\pi}
	\end{equation}
and
	\begin{equation}
		\hat{h}^{\mu\nu}E_{\mu\nu\rho_{1}\sigma_{1}\rho_{2}\sigma_{2}\rho_{3}\sigma_{3}}\bar{\nabla}^{\rho_{1}}\bar{\nabla}^{\sigma_{1}}\hat{\pi}\bar{\nabla}^{\rho_{2}}\bar{\nabla}^{\sigma_{2}}\hat{\pi}\bar{\nabla}^{\rho_{2}}\bar{\nabla}^{\sigma_{3}}\hat{\pi},
	\end{equation}
are
	\begin{eqnarray}
	0 & = & -2\bar{\nabla}^{\mu}\left(E_{\mu\nu\rho_{1}\sigma_{1}\rho_{2}\sigma_{2}}\bar{\nabla}^{\nu}\hat{\pi}\right)\bar{\nabla}^{\rho_{1}}\bar{\nabla}^{\sigma_{1}}\hat{\pi}\bar{\nabla}^{\rho_{2}}\bar{\nabla}^{\sigma_{2}}\hat{\pi}+4E_{\mu\nu\rho_{1}\sigma_{1}\rho_{2}\sigma_{2}}\bar{R}^{\rho_{1}\mu\sigma_{1}\lambda}\bar{\nabla}_{\lambda}\hat{\pi}\bar{\nabla}^{\nu}\hat{\pi}\bar{\nabla}^{\rho_{2}}\bar{\nabla}^{\sigma_{2}}\hat{\pi}\nonumber \\
	 &  & +\left(\bar{\nabla}^{\sigma_{1}}E_{\mu\nu\rho_{1}\sigma_{1}\rho_{2}\sigma_{2}}\bar{\nabla}^{\mu}\hat{\pi}+2E_{\mu\nu\rho_{1}\sigma_{1}\rho_{2}\sigma_{2}}\bar{\nabla}^{\sigma_{1}}\bar{\nabla}^{\mu}\hat{\pi}\right)\bar{\nabla}^{\nu}\hat{\pi}\bar{R}^{\rho_{1}\rho_{2}\sigma_{2}\lambda}\bar{\nabla}_{\lambda}\hat{\pi}\nonumber \\
	 &  & +2\bar{\nabla}^{\rho_{1}}\left(\bar{\nabla}^{\sigma_{1}}E_{\mu\nu\rho_{1}\sigma_{1}\rho_{2}\sigma_{2}}\bar{\nabla}^{\mu}\hat{\pi}\bar{\nabla}^{\nu}\hat{\pi}\right)\bar{\nabla}^{\rho_{2}}\bar{\nabla}^{\sigma_{2}}\hat{\pi}+4\bar{\nabla}^{\rho_{1}}\left(E_{\mu\nu\rho_{1}\sigma_{1}\rho_{2}\sigma_{2}}\bar{\nabla}^{\nu}\hat{\pi}\right)\bar{\nabla}^{\sigma_{1}}\bar{\nabla}^{\mu}\hat{\pi}\bar{\nabla}^{\rho_{2}}\bar{\nabla}^{\sigma_{2}}\hat{\pi}\nonumber \\
	 &  & +\bar{\nabla}^{\rho_{1}}\left(E_{\mu\nu\rho_{1}\sigma_{1}\rho_{2}\sigma_{2}}\bar{\nabla}^{\mu}\hat{\pi}\bar{\nabla}^{\nu}\hat{\pi}\bar{R}^{\sigma_{1}\sigma_{2}\rho_{2}\lambda}\bar{\nabla}_{\lambda}\hat{\pi}\right),
	\end{eqnarray}
and
	\begin{equation}
		0=\hat{h}^{\mu\nu}E_{\mu\nu\rho_{1}\sigma_{1}\rho_{2}\sigma_{2}\rho_{3}\sigma_{3}}\left[\bar{\nabla}^{\rho_{1}}\left(\bar{R}^{\sigma_{1}\sigma_{2}\rho_{2}\lambda}\bar{\nabla}_{\lambda}\hat{\pi}\right)\bar{\nabla}^{\rho_{2}}\bar{\nabla}^{\sigma_{3}}\hat{\pi}+\bar{R}^{\sigma_{1}\sigma_{2}\rho_{2}\lambda}\bar{R}^{\rho_{1}\rho_{2}\sigma_{3}\tau}\bar{\nabla}_{\lambda}\hat{\pi}\bar{\nabla}_{\tau}\hat{\pi}\right],
	\end{equation}
respectively, which also contain no higher derivatives of $\hat{\pi}$.
Terms in $\mathcal{L}_4$ are just special cases of the above.

\section{USEFUL INTEGRATION BY PARTS} \label{sec:ibp}

In the case of a flat fiducial metric, the self-interactions of the
helicity-0 mode $\pi$ exactly vanish because they become total
derivatives. In the case of a curved fiducial metric, instead we have,
at the quadratic order
	\begin{equation}
		\left(\bar{\square}\pi\right)^{2}-\bar{\nabla}_{\mu}\bar{\nabla}_{\nu}\pi\bar{\nabla}^{\mu}\bar{\nabla}^{\nu}\pi\simeq\bar{R}_{\mu\nu}\bar{\nabla}^{\mu}\pi\bar{\nabla}^{\nu}\pi, \label{ibp2}
	\end{equation}
at the cubic order
	\begin{eqnarray}
	 &  & \left(\bar{\square}\pi\right)^{3}-3\bar{\square}\pi\bar{\nabla}_{\mu}\bar{\nabla}_{\nu}\pi\bar{\nabla}^{\mu}\bar{\nabla}^{\nu}\pi+2\bar{\nabla}_{\mu}\bar{\nabla}^{\nu}\pi\bar{\nabla}_{\nu}\bar{\nabla}^{\rho}\pi\bar{\nabla}_{\rho}\bar{\nabla}^{\mu}\pi\nonumber \\
	 & \simeq & 2\bar{R}_{\mu\nu}\bar{\nabla}^{\nu}\pi\left(\bar{\nabla}^{\mu}\pi\bar{\square}\pi-\bar{\nabla}^{\mu}\bar{\nabla}_{\rho}\pi\bar{\nabla}^{\rho}\pi\right)-2\bar{R}_{\mu\rho\nu\sigma}\bar{\nabla}^{\rho}\pi\bar{\nabla}^{\sigma}\pi\bar{\nabla}^{\mu}\bar{\nabla}^{\nu}\pi,\label{ibp3}
	\end{eqnarray}
and at the quartic order
	\begin{eqnarray}
	 &  & \left(\bar{\square}\pi\right)^{4}-6\left(\bar{\square}\pi\right)^{2}\bar{\nabla}_{\mu}\bar{\nabla}_{\nu}\pi\bar{\nabla}^{\mu}\bar{\nabla}^{\nu}\pi+8\bar{\square}\pi\bar{\nabla}^{\mu}\bar{\nabla}_{\nu}\pi\bar{\nabla}^{\nu}\bar{\nabla}_{\rho}\pi\bar{\nabla}^{\rho}\bar{\nabla}_{\mu}\pi\nonumber \\
	 &  & +3\left(\bar{\nabla}_{\mu}\bar{\nabla}_{\nu}\pi\bar{\nabla}^{\mu}\bar{\nabla}^{\nu}\pi\right)^{2}-6\bar{\nabla}^{\mu}\bar{\nabla}_{\nu}\pi\bar{\nabla}^{\nu}\bar{\nabla}_{\rho}\pi\bar{\nabla}^{\rho}\bar{\nabla}_{\sigma}\pi\bar{\nabla}^{\sigma}\bar{\nabla}_{\mu}\pi\nonumber \\
	 & \simeq & 3\bar{R}_{\mu\nu}\bar{\nabla}^{\mu}\pi\bar{\nabla}^{\nu}\pi\left[\left(\bar{\square}\pi\right)^{2}-\bar{\nabla}_{\rho}\bar{\nabla}_{\sigma}\pi\bar{\nabla}^{\rho}\bar{\nabla}^{\sigma}\pi\right]-6\bar{R}_{\mu\nu}\bar{\nabla}^{\alpha}\pi\bar{\nabla}^{\mu}\pi\left(\bar{\nabla}^{\nu}\bar{\nabla}_{\alpha}\pi\bar{\square}\pi-\bar{\nabla}^{\rho}\bar{\nabla}_{\alpha}\pi\bar{\nabla}^{\nu}\bar{\nabla}_{\rho}\pi\right)\nonumber \\
	 &  & -6\bar{R}_{\mu\rho\nu\sigma}\left[\bar{\nabla}^{\mu}\pi\bar{\nabla}^{\nu}\pi\left(\bar{\square}\pi\bar{\nabla}^{\rho}\bar{\nabla}^{\sigma}\pi-\bar{\nabla}^{\lambda}\bar{\nabla}^{\rho}\pi\bar{\nabla}^{\sigma}\bar{\nabla}_{\lambda}\pi\right)-\bar{\nabla}^{\alpha}\pi\bar{\nabla}^{\mu}\pi\bar{\nabla}^{\nu}\bar{\nabla}_{\alpha}\pi\bar{\nabla}^{\rho}\bar{\nabla}^{\sigma}\pi\right].\label{ibp4}
	\end{eqnarray}

\end{document}